\documentclass[twocolumn,aps,prb,superscriptaddress,nofootinbib]{revtex4-1}
\usepackage{graphicx}
\usepackage{times}
\usepackage{bm}
\usepackage[linktocpage=true,colorlinks=true,urlcolor=blue]{hyperref}
\usepackage{pgfplots}
\usepackage{amsfonts}
\usepackage{amsmath}
\usepackage{xcolor}
\usepackage{braket}
\usepackage{relsize}
\usetikzlibrary{arrows}
\usepackage{enumerate}
\usetikzlibrary[patterns] 
\usetikzlibrary[snakes] 
\usetikzlibrary[shapes] 
\usepackage{pgfplots}
\usepackage{newfloat}
\DeclareFloatingEnvironment[name={Supplementary Figure}]{suppfigure}

\graphicspath{ {Figures2/} }

\begin{document}

\title{Zero-energy pinning of topologically-trivial bound states in multi-band semiconductor-superconductor nanowires}

\author{Benjamin D. Woods}
\affiliation{Department of Physics and Astronomy, West Virginia University, Morgantown, WV 26506, USA}
\author{Jun Chen}
\affiliation{Department of Physics and Astronomy, University of Pittsburgh, Pittsburgh, PA 15260, USA}
\affiliation{Department of Electrical and Computer Engineering, University of Pittsburgh, Pittsburgh, PA 15261, USA}
\author{Sergey M. Frolov}
\affiliation{Department of Physics and Astronomy, University of Pittsburgh, Pittsburgh, PA 15260, USA}
\author{Tudor D. Stanescu}
\affiliation{Department of Physics and Astronomy, West Virginia University, Morgantown, WV 26506, USA}

\begin{abstract}
Recent tunneling conductance measurements on semiconductor-superconductor nanowires find zero-bias peaks to be ubiquitous across wide ranges of chemical potential and Zeeman energy \cite{Chen2019}. 
 Motivated by this, we demonstrate that topologically-trivial Andreev abound states (ABSs) pinned near zero energy are produced rather generically in inhomogeneous systems with multi-band occupancy in the presence of inter-band coupling. We first  investigate the  inter-band coupling mechanism responsible for the pinning within a multi-band 1D toy model, then we confirm the findings using a 3D Schr\"{o}dinger-Poisson approach  that incorporates the geometric and electrostatic details of the actual device. Our analysis shows that level-repulsion generated by inter-band coupling can lead  to a rather spectacular pinning of the lowest-energy mode near zero energy in systems (or regions) characterized by very-short length scales ($\sim100~$nm).
 We show that level repulsion between the  lowest energy levels can mimic the gap opening feature (simultaneous with the emergence of a near-zero energy mode) predicted to occur in Majorana hybrid systems. We also demonstrate that nearly-zero bias differential conductance features exhibiting particle-hole asymmetry are due to the presence of (topologically-trivial) ABSs pinned near zero-energy by level repulsion, not to  Majorana zero modes, quasi-Majoranas, or any other low-energy mode that involves (partially) separated Majorana bound states. 
 Our findings demonstrate the importance of understanding in detail multi-band physics and electrostatic effects in the context of the ongoing search for Majorana modes  in semiconductor-superconductor heterostructures.
\end{abstract}

\maketitle

\section{Introduction}
Majorana zero-energy modes (MZMs) are promising candidates for qubits in future fault-tolerant topological quantum computers.\cite{Bravyi2002,Nayak2008,Stanescu2017,Lahtinen2017} These modes have been predicted to emerge as zero-energy bound states in proximity-coupled semiconductor-superconductor (SM-SC) hybrid structures in the presence of strong spin-orbit coupling and finite Zeeman splitting.\cite{SauPRL2010a,Alicea2010,Oreg2010,Lutchyn2010,Sau2010} Despite encouraging results and significant experimental progress,\cite{Mourik2012,Deng2012,Das2012,Rokhinson2012,Churchill2013,Finck2013,Albrecht2016,Deng2016,Zhang2017,Chen2017,Nichele2017,Zhang2018} 
a persistent concern within the field has been whether the entire set of conditions necessary for the topological superconducting phase and the associated MZMs is actually realized in the laboratory and, consequently, whether the observed experimental features (e.g., the presence of zero-bias peaks in the differential conductance at finite magnetic field) should be attributed to the sought after MZMs.
One reason behind this concern is that several of the predicted key features associated with the presence of  MZMs have yet to be demonstrated, e.g., the correlated conductance signatures at the opposite ends of a short wire generated by the energy splitting oscillations due to the partial overlap of the two MZMs.\cite{DSarma2012} Another important reason is the relative ease with which zero-bias conductance peaks (ZBCPs) are ubiquitously observed in tunneling measurements on proximitized SM nanowires,
as if the system conspires to make the topological condition  $\Gamma >\Gamma_c=\sqrt{\mu^2+\Delta^2}$ easily realizable by `keeping' the chemical potential close to the bottom of one of the confinement-induced sub-bands (from now on simply called {\em bands}). Here, $\Gamma$ is the Zeeman field, $\Gamma_c$ is the critical field associated with the topological quantum phase transition (TQPT), $\Delta$ the induced SC pairing potential, and $\mu$ the chemical potential (relative to the top occupied sub-band).
The concern is amplified by works demonstrating the possibility of having topologically-trivial low-energy Andreev bound states (ABSs)  that mimic the phenomenology of topologically-protected MZMs in wires with disorder,\cite{Bagrets2012,Liu2012,DeGottardi2013,Rainis2013,Adagideli2014} non-uniform effective parameters,\cite{Kells2012,Chevallier2012,Roy2013,SanJose2013,Ojanen2013,Stanescu2014,Cayao2015,Klinovaja2015,SanJose2016,Fleckenstein2018} and weak antilocalizaton\cite{Pikulin2012}, or in wires coupled to a quantum dot.\cite{Prada2012,Liu2017a,Ptok2017}

Understanding the possible mechanisms that cause the pinning of ABSs near zero energy is important for i) distinguishing topological MZMs from non-topological ABSs with similar phenomenology and ii) estimating the relative sizes of the parameter subspaces consistent with the formation of MZMs and low-energy ABSs, respectively, and explaining the apparent ubiquitousness of zero-bias conductance peaks in tunneling experiments. Ultimately, these efforts are meant to guide the engineering of SM-SC hybrid structures toward the reliable, controlled realization of the conditions consistent with the emergence of MZMs. In general, low-energy ABSs can emerge at Zeeman fields below the critical value $\Gamma_c$ associated with the TQPT when the system is not homogeneous.
The inhomogeneity is generated either by the structure of the hybrid device, e.g., the presence of (multiple) electrostatic gates, the partial covering of the SM wire with superconducting material, and the presence of multiple tunnel contacts,\cite{Stanescu2018a,Huang2018,Grivnin2018} or by disorder at surfaces and interfaces, e.g., disorder at the surface of the SC film and  spatial fluctuations of the work function difference between the SC and the SM wire.\cite{Woods2018} The most direct consequence of having such sources of inhomogeneity is an effective potential that varies along the wire, which, in turn, gives rise to low-energy ABSs.\cite{Stanescu2018b,Moore2018}  

One possible explanation for the emergence of ABSs pinned near zero energy over an extended range of parameters\cite{Moore2018a,Reeg2018b,Vuik2019,Stanescu2018b}  (e.g., applied magnetic field)
is the partial spatial separation of the two Majorana modes \cite{Moore2018a,Stanescu2018b}  (also known as quasi-Majorana states\cite{Vuik2019}) that constitute the Andreev bound state, which was dubbed a partially-separated ABS (ps-ABS). This partial separation mechanism is quite generic, being responsible for the emergence of low-energy ABSs in various types of non-homogeneous systems, e.g., in the presence of  smooth confinement, potential wells/hills, or quantum dots coupled to a proximitized wire.
Note, however,  that this mechanism was studied theoretically based (almost exclusively) on i) single-band tight-binding models and ii) toy models for the potential profile containing largely arbitrary parameter values. Most importantly, the partial-separation mechanism does not explain the ubiquity of zero-bias features observed experimentally.\cite{Chen2019} Therefore, exploring the possibility that other mechanisms may also be responsible for the collapse of ABSs toward zero energy in systems with multi-band occupancy represents a critical outstanding task. 
In addition, the potential profiles (and other position-dependent system parameters) should be calculated based on the properties of actual hybrid devices, rather than postulated based on arbitrary assumptions. For example, in systems with soft confinement the collapse (and `sticking') to zero-energy of a ps-ABS localized near the end of the wire is controlled by the slope of the potential.\cite{Kells2012,Vuik2019,Stanescu2018b} In turn, estimating this slope requires solving a challenging 3D Schr\"{o}dinger-Poisson problem that takes into account the geometry of the heterostructure and the applied gate potentials.\cite{Woods2018}  
Without explicitly solving this type of problem, it is difficult (if not impossible) to estimate if the conditions necessary for the emergence of low-energy ABSs (through either the partial separation mechanism, or the newly-proposed inter-band coupling mechanism) are generically satisfied, somewhat likely, or nearly impossible.

While our conclusions are general, we focus on a recent tunneling conductance experiment on InSb/NbTiN  hybrid structures, in which low-energy features similar to those predicted theoretically were observed over a considerable parameter range (i.e., tunnel barrier and back gate potentials, magnetic fields, etc.).\cite{Chen2019} Are they generated by MZMs, ps-ABSs (i.e., quasi-Majoranas), or some other (topologically-trivial) low-energy ABSs? The experiment provides some useful hints. First, we note that the characteristic length scales associated with the structural inhomogeneity of the device are small: an uncovered region
of about $100~$ nm (corresponding to the tunneling gate region) and a covered region defined by a bottom gate of about $200~$nm.\cite{Chen2019} These small length scales pose serious difficulties to the partial-separation scenario.\cite{Stanescu2018b} Second, the single-band model calculations predict that the low-energy ABSs should be fairly well separated in energy from the bulk states (on the scale of the induced SC gap); by contrast, the experiment shows a relatively crowded sub-gap spectrum.

In this work we show that multi-band occupancy characterizes the hetero-structure studied in the recent tunneling conductance experiment\cite{Chen2019} over the whole relevant range of control parameters. Furthermore, we demonstrate that, in general, hybrid systems with multi-band occupancy host topologically trivial ABSs that can be pinned near zero energy as a result of a mechanism that involves the coupling of two or more confinement-induced low-energy bands. Band repulsion resulting from this coupling pins the lowest energy state near zero energy over a significant range of control parameters (e.g., Zeeman field).  We emphasize that multi-band occupancy (an ingredient that is not included in the vast majority of the theoretical work on Majorana hybrid structures) is crucial for this mechanism to be active. 
The resulting (topologically trivial) ABSs are characterized by Majorana modes that are {\em not} separated spatially. Consequently, the characteristic length scales associated with the collapse and pinning to zero energy of the ABSs generated by this mechanism are significantly smaller (e.g. on the order of the nanowire diameter of $100$ nm) than those required for the formation of a ps-ABS. In general, in addition to the partial-separation mechanism discussed extensively in the literature, the inter-band coupling mechanism should be viewed as an alternative path for generating low-energy ABSs.  This mechanism becomes dominant in systems with multi-band occupancy and short-range inhomogeneities. We show that this inter-band coupling mechanism is capable of explaining the features observed in the experimental data reported in Ref \onlinecite{Chen2019}. 
In addition, we find that, unlike (partially) separated Majorana modes, the ABS modes generated by the inter-band coupling mechanism retain their particle {\em or} hole character down to zero energy (except for a few discrete points). Consequently, in the presence of dissipation, the nearly-zero energy conductance features generated by these ABSs can break particle-hole symmetry. We conclude that the observation of \textit{nearly}-zero differential conductance peaks that break particle-hole symmetry, which is inconsistent with the presence of MZMs, quasi-Majoranas, or any other low-energy modes that involve (partially) separated Majorana bound states, should be attributed to (topologically-trivial) ABSs pinned near zero-energy by level repulsion. 

To incorporate the details of the electrostatic environment characterizing the experimental device, we perform 3D Schr\"{o}dinger-Poisson calculations using an efficient approach developed earlier.\cite{Woods2018} 
We match the geometry of the device (gate sizes, material parameters, superconductor geometry, etc.) in an attempt to be as close as possible to the relevant parameter regime. Within this approach, we demonstrate that inter-band coupling is a direct (and necessary) consequence of the inhomogeneous electrostatic potential \textit{along} the wire.
Moreover, inter-band coupling is expected to be a generic feature at interfaces between regions with different electrostatic environments. 
Note that the present calculation does not include disorder, which is expected to induce additional interband coupling. 
Nonetheless, these results emphasize the importance of being able to perform 3D Schr\"{o}dinger-Poisson calculations, rather than assuming translation invariance along the wire. Since our realistic modeling predicts i) multi-band occupancy over the whole range of experimentally-relevant control parameters and ii) strong inter-band coupling, we conclude that the emergence of low-energy ps-ABSs generated by the inter-band coupling mechanism is quite generic, in agreement with the ubiquity of zero-bias conductance peaks observed experimentally. 

The remainder of this work is organized as follows. In Sec. \ref{Toy}, we present a toy model that illustrates the basic principle behind the inter-band coupling mechanism. We introduce the key ideas associated with inter-band coupling in SM-SC hybrid structures and explore multi-band effects in both homogeneous wires (Sec. \ref{Homo}) and inhomogeneous systems (Sec. \ref{Inhomo}). A detailed 3D model of the device is described in Sec. \ref{3D theory} and the corresponding results, which show explicitly that inter-band coupling can pin ABSs near zero energy, are presented in Sec. \ref{3D Results}. Finally, in Sec. \ref{Discussion} we discuss the relevance of our findings and suggest ways to lessen the chance of ABSs emerging as a result of the inter-band coupling mechanism.

\section{Toy model} \label{Toy}
To illustrate the main ideas underlying the emergence of low-energy ABSs within the inter-band coupling mechanism and to emphasize the main differences between this multi-band scenario and the partial separation mechanism responsible for the formation of ps-ABSs
in single-band models, we  first consider a multi-band tight-binding toy model that captures the essential aspects of  multi-band physics in hybrid structures. 
We emphasize that the ABSs emerging within the inter-band coupling mechanism consist of two overlapping (i.e. non-separated) Majorana modes, yet they still `stick' near zero energy over a wide range of magnetic field. 
This is in stark contrast to single band scenarios (or, in general, models that neglect inter-band coupling), which predict either ps-ABSs (consisting of two Majorana modes with significant separation in space \cite{Vuik2019,Moore2018}) that stick near zero-energy, or ``plain'' ABSs (composed of overlapping Majorana modes) that can only cross zero-energy, without ``sticking''. 
Note that the presence of various inter-band coupling terms in the toy model can be fully justified based on the 3D self-consistent calculations presented in Sec. \ref{3Dmodel}. However, the toy model has the major advantage that, due to its (relative) simplicity,  it makes the physics behind the  inter-band coupling mechanism  more transparent. Specifically, we consider the following 1D nearest neighbor multi-band Hamiltonian describing a finite SM nanowire weakly coupled to an s-wave superconductor:
\begin{equation}
\begin{aligned}
H &= \sum\limits_i^{N-1} \sum\limits_{\alpha,\beta} \sum\limits_{\sigma}
	t_i^{\alpha\beta} c_{i\alpha\sigma}^\dagger c_{i+1,\beta\sigma} + h.c. \\
	&+ \sum\limits_{i\alpha} \sum\limits_{\sigma\sigma^\prime}
	\left[\left(V_{i\alpha} - \mu_\alpha - 2t_i^{\alpha\alpha}\right)\delta_{\sigma,\sigma^\prime} 
	+ \Gamma \left(\sigma_x\right)_{\sigma\sigma^\prime}\right]
	 c_{i\alpha\sigma}^\dagger c_{i\alpha\sigma^\prime} \\
	 &+\sum\limits_i^{N-1} \sum\limits_{\alpha,\beta} \sum\limits_{\sigma\sigma^\prime}
	 \widetilde{\alpha}_i^{\alpha\beta }
	 c_{i\alpha\sigma}^\dagger \left(i\sigma_y\right)_{\sigma\sigma^\prime}
	  c_{i+1,\beta\sigma^\prime} + h.c. \\
	 &+ \sum\limits_i \sum\limits_{\alpha\beta} \sum\limits_{\sigma\sigma^\prime}
	 \left(\alpha_{T}\right)_{i}^{\alpha\beta}  
	 c_{i\alpha\sigma}^\dagger \left(i\sigma_x\right)_{\sigma\sigma^\prime}
	  c_{i\beta\sigma^\prime} + h.c. \\
	 &+ \sum\limits_{i\alpha\beta} \Delta_i^{\alpha\beta} 
	 c_{i\alpha\uparrow}^\dagger c_{i\beta\downarrow}^\dagger + h.c.,
\end{aligned}  \label{Htoy}
\end{equation}
where $c_{i\alpha\sigma}$ annihilates an electron on the $i^{th}$ site with $\alpha$ and $\sigma$ being band and spin indices, respectively, $t_i^{\alpha\beta}$ is a spin-conserving hopping matrix element, $\Gamma$ is the (half) Zeeman splitting due an external magnetic field, $V_{i\alpha}$ is the effective potential 
of the $\alpha$ band, $\Delta_i^{\alpha\beta}$ is a superconducting pairing matrix element, while $\widetilde{\alpha}_i^{\alpha\beta }$ and $\left(\alpha_{T}\right)_i^{\alpha\beta }$ are longitudinal and transverse spin-orbit matrix elements, respectively.
The parameter $\mu_\alpha = \mu - \epsilon_\alpha$, where $\mu$ is the chemical potential and $\epsilon_\alpha$ is the energy of the $\alpha$ band at $k=0$ (in a long, uniform wire), represents the chemical potential relative to the bottom of the band. Note that subtracting the quantity $2t_i^{\alpha\alpha}$ from the on-site energy ensures that the bottom of the $\alpha$ band is at $\epsilon_\alpha$.
Finally, $\sigma_i$, with $i=x,y,z$, are Pauli spin matrices. 
 We note that $\widetilde{\alpha}_i^{\alpha\beta } = \alpha_i^{\alpha\beta}/2a$, where $a$ is the lattice constant of the 1D lattice describing the wire, and $\alpha_i^{\alpha\beta}$ has units of energy times length (i.e. the typical units for the spin-orbit coupling constant). By contrast, $\left(\alpha_{T}\right)_i^{\alpha\beta }$ does not scale with the lattice spacing, as it models transverse spin orbit coupling between various orbitals delocalized across the transverse section of the wire. The multi-band model (\ref{Htoy}) reduces to the `standard' single-band model used in the literature if all the matrices are diagonal,  $\mathcal{O}^{\alpha\beta} = 0$ for $\alpha \neq \beta$ (i.e., there is no inter-band coupling), and one focuses on the top occupied band. 
We emphasize that the multi-band nature of the model (which involves inter-band coupling as an essential ingredient)  introduces new physics that is relevant to understanding many of the features observed in the current experiments on semiconductor-superconductor hybrid structures, as we demonstrate below. 

The parameter values used in the numerical calculations are loosely based on the (known) parameters for a typical SM-SC structures (e.g., InAs nanowires proximitized with Al) and take into consideration certain symmetry constraints, as discussed below. 
However, the main point of this section is not to provide quantitative predictions (e.g., to fit specific experimental results), but rather to reveal the role of inter-band coupling in generating low-energy ABSs pinned near zero energy. By contrast, the inter-band couplings obtained within the full 3D calculation of Sec. \ref{3Dmodel} are not arbitrary, being determined by the evolution of the transverse profile of the orbitals associated with different confinement-induced bands (determined self-consistently) as one moves along the wire.

\begin{figure}[t]
\begin{center}
\includegraphics[width=0.48\textwidth]{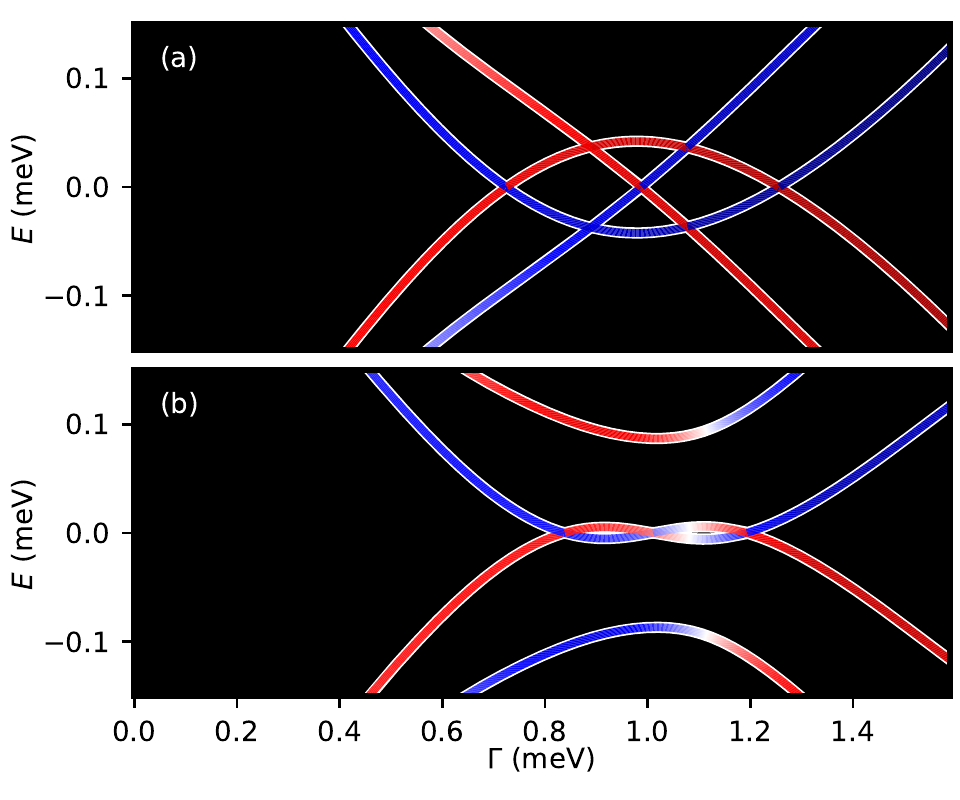}
\end{center}
\vspace{-5mm}
\caption{Spectrum of a two band model with the chemical potential placed between the two bands for a wire of length $L=250~$nm. The model parameters are given by $m^* = 0.03$, $\alpha_{1,1}=\alpha_{2,2}=500$ meV \AA, $\Delta_{1,1}=\Delta_{2,2} = 0.35$ meV, $\mu_1 = 0.28$ meV, and $\mu_2 = -0.35$ meV, where $\mu_\alpha$ is defined with respect to the bottom of the corresponding band. (a) No inter-band coupling. (b) Inter-band coupling defined by $\Delta_{1,2} = -\Delta_{2,1} = 0.21$ meV and $(\alpha_T)_{1,2} = 0.19$ meV. Red and blue lines indicate (dominant) particle and hole weights, respectively, while white denotes an equal particle-hole mixture.}
\label{FIG1}
\vspace{-3mm}
\end{figure}

\subsection{Homogeneous wires} \label{Homo}
First, let us consider a short homogeneous system with position-independent effective potential $V_{i\alpha}=0$ (for all bands). In addition, we require the Hamiltonian to respect inversion symmetry (see Appendix \ref{App1}). 
Note that in the homogeneous case, the near-zero energy states are ABSs consisting of strongly overlapping Majorana modes. We focus on the effects of the inter-band coupling on these low-energy states. More specifically, let us consider a short wire of length $L = 250~$nm.
We describe the wire using a two-band model and assume that the chemical potential lays between the (bottoms of the) two bands, $\epsilon_1 < \mu < \epsilon_2$.  
The corresponding low-energy spectrum is shown in Fig. \ref{FIG1}.  If the two bands are decoupled, see Fig. \ref{FIG1}(a), the empty and occupied bands generate particle- (red) and hole-type (blue) states coming toward zero energy from above as the Zeeman field increases from zero. Note that in this discussion we focus on the positive energy sector, but the spectrum is particle-hole symmetric, as clearly shown in Fig. \ref{FIG1}.  The particle mode has one zero energy crossing near $\Gamma\approx 0.95~$meV and becomes a hole mode (blue) at larger fields (within the positive energy sector). The hole band exhibits one ``oscillation'', with two zero-energy crossings (at $\Gamma\approx 0.67~$meV and $\Gamma\approx 1.3~$meV, respectively) within the relevant Zeeman field range.
The energy splittings of the low-energy modes (which can be viewed as being induced by the strong overlap of the Majorana components of these modes) are large, i.e. comparable to the induced pairing potential. This is a finite-size effect generated by the short length of the wire, which does not allow the separation of the Majorana components. Note that, formally, the topological condition $\Gamma >\sqrt{\mu_\alpha^2+\Delta_{\alpha\alpha}^2}$ is satisfied for both bands when $\Gamma>0.5~$meV.  
Next, upon introducing an inter-band coupling through the off-diagonal pairing and the transverse spin-orbit coupling, the two bands hybridize and generate a mode that sticks to zero energy over a finite range of Zeeman fields, as shown in  Fig. \ref{FIG1}(b) . Note that the lowest-energy mode undergoes two oscillations about zero energy with amplitude significantly lower than the energy splittings characterizing the decoupled system. We conclude that level repulsion induced by inter-band mixing can generate a low-energy mode that sticks to zero-energy over a finite range of Zeeman field. 

\begin{figure}[t]
\begin{center}
\includegraphics[width=0.48\textwidth]{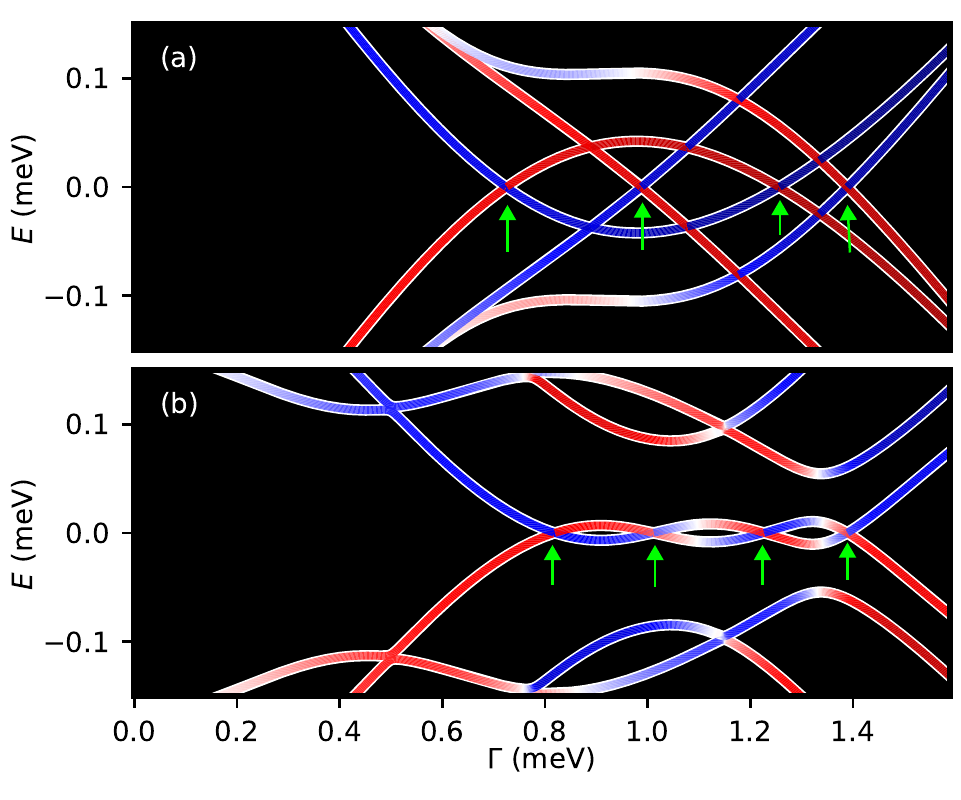}
\end{center}
\vspace{-5mm}
\caption{Spectrum of a three band model with the same parameters for the first two bands as in Fig. \ref{FIG1},  $\mu_3=-0.6$ meV, and $\Delta_{3,3} = 0.35$ meV.  (a) No inter-band coupling. (b) Inter-band coupling defined by  $\Delta_{2,3} = \Delta_{3,2} = 0.2~$meV, $(\alpha_T)_{1,3} = 0.05~$meV,  in addition to the parameters defined in  Fig. \ref{FIG1}(b).}
\label{FIG2}
\vspace{-3mm}
\end{figure}

To strengthen this conclusion, we consider a three-band model of the short wire ($L=250~$nm), both with and without inter-band coupling. The corresponding spectra are shown in Fig. \ref{FIG2}. The parameters of the model corresponding to the first two bands are the same as in Fig. \ref{FIG1}, while the parameters associated with the  third band are provided in the figure caption. Note that, again, in the presence of inter-band coupling, level repulsion pushes one mode toward zero energy over a significant range of Zeeman field (about $0.6~$meV). The near-zero-energy mode is characterized by three low-amplitude oscillations, which may appear in tunneling spectroscopy as a robust ZBCP (without splitting) due to broadening and finite energy resolution. Furthermore, even the splitting of the low-energy mode away from zero energy for $\Gamma > 1.4~$meV may not be observable in practice, if the the superconducting gap of the parent superconductor collapses at comparable values of the magnetic field.
We conclude that the  inter-band coupling mechanism illustrated in these examples can generate low-energy states with local signatures similar to those of topologically-protected MZMs (e.g., a robust ZBCP), despite the wire being very short (i.e. being incapable to support two well-separated Majorana modes localized at the opposite ends of the system). 

\begin{figure}[t]
\begin{center}
\includegraphics[width=0.48\textwidth]{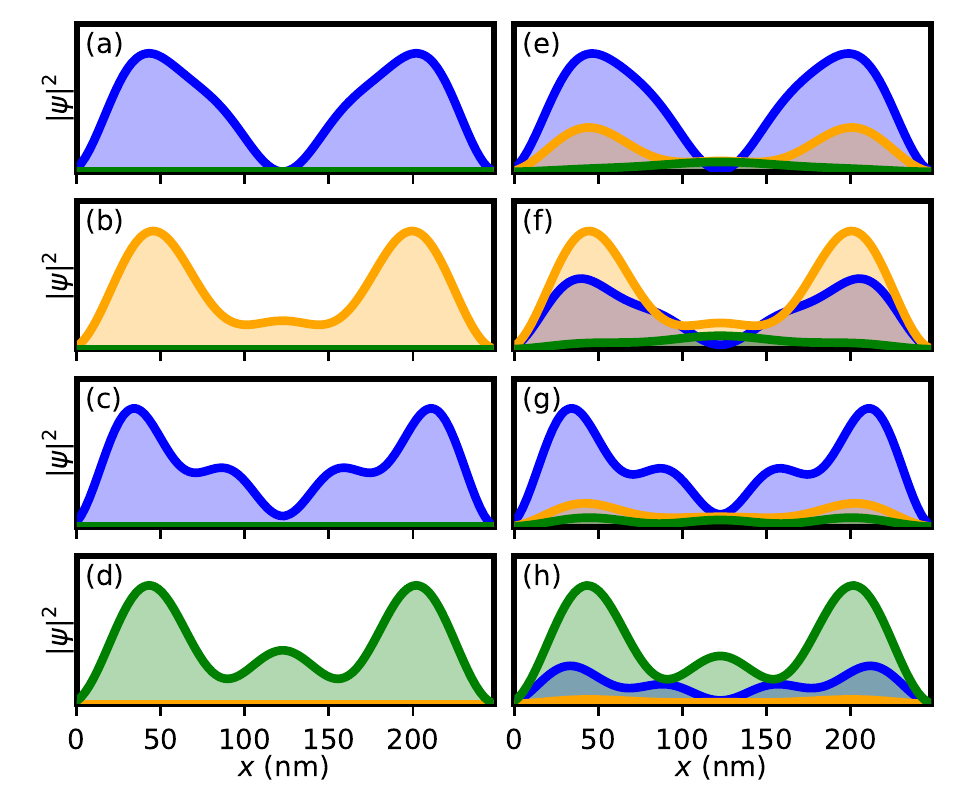}
\end{center}
\vspace{-5mm}
\caption{Wave function amplitudes of the zero energy states indicated by green arrows in Fig. (\ref{FIG2}). Panels (a)-(d) and (e)-(h) correspond to Fig. \ref{FIG2}(a) and Fig. \ref{FIG2}(b), respectively. The three colors correspond to different band contributions. The spin and particle-hole degrees of freedom are summed over for each lattice site. 
}
\label{FIG3}
\vspace{-3mm}
\end{figure}

To gain further insight, we calculate the position-dependence of the wave function amplitude for  the states corresponding to the zero energy crossings in Fig. \ref{FIG2} (see the green arrows). The results are shown in Fig. \ref{FIG3}. Note that in the decoupled-band case [panels (a) - (d) in Fig. \ref{FIG3}], each state is composed  of a single band component. By contrast, in the presence of  inter-band coupling, the zero energy states are composed of a mixture of the three bands [panels (e) - (h)]. Note, however, that the separation between the main wave function peaks is similar in the two cases, suggesting that the collapse to zero-energy of the low-energy mode in the band-coupled system  is not the result of the component Majorana modes becoming spatially separated. 
In fact, the explicit calculation of the corresponding Majorana components (see Appendix \ref{App2} for a technical definition) shows that they have nonzero amplitude throughout the entire wire and cannot be identified with the main peaks of the ABS wave function. We emphasize that previous studies using single-band models found the pinning to zero of a low-energy mode to be  necessarily associated with the (partial) separation of the Majorana modes.\cite{Moore2018,Stanescu2018b}
By contrast, ABSs generated by the inter-band coupling mechanism are not characterized by separated  Majorana modes.  
We note that inter-band level repulsion in homogeneous systems has been previously explored,\cite{Moor2018} being attributed to spin-orbit coupling. However, it should be pointed out that inter-band coupling can be more general, e.g., it can involve the (induced) superconducting pairing potential, as shown here. This effect can be naturally understood as a proximity-induced coupling of the confinement-induced bands and is expected to be significant in the strong coupling limit.\cite{Stanescu2017a} More importantly, homogeneous systems represent a rather ideal limit which may not be easily realizable in practice. A more interesting (and potentially relevant) situation involves inter-band mixing caused by an inhomogeneous electrostatic potential, which we address in Sec. \ref{Inhomo}. 

\begin{figure}[t]
\begin{center}
\includegraphics[width=0.48\textwidth]{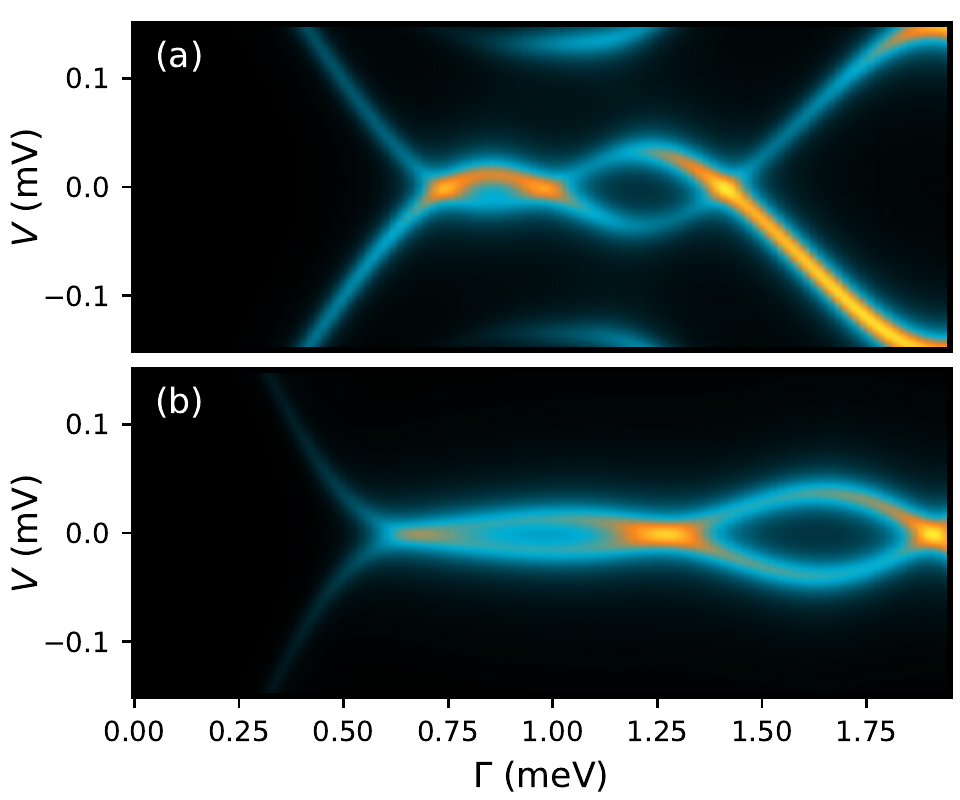}
\end{center}
\vspace{-5mm}
\caption{Differential conductance as function of Zeeman field and bias voltage in the presence of dissipation for a hybrid system that supports (a) an ABS generated by the inter-band coupling mechanism and (b) Majorana bound states in a finite wire.  Note that the ABS generates a particle-hole asymmetric zero bias peak, in contrast with the MBS signature, which is particle-hole symmetric. The system parameter for panel (a) are the same as those in Fig. \ref{FIG1} except $\mu_1 = 0.2~$meV and $\left(\alpha_T\right)_{1,2} = 0.2~$meV. In panel (b) we have a wire of length $780~$nm described by a single-band model with $\mu=0.25~$meV,  $m^* = 0.03$, $\alpha_L=500~$meV \AA, and $\Delta = 0.35~$meV. Dissipation was modeled as an imaginary contribution of magnitude $\eta=0.015~$meV.}
\label{FIGx}
\vspace{-3mm}
\end{figure}

Before closing this section, we want to emphasize a fundamental difference between the ABSs generated by the inter-band coupling mechanism described here and the (partially) separated Majorana modes that emerge in a finite wire upon satisfying the `topological condition' $\Gamma >\Gamma_c$. While the Majorana modes  represent (almost) equal mixtures of particles and holes (hence, they appear as white lines in our color code representation, see, for example, Fig. \ref{FIG4}), the ABS modes retain their particle {\em or} hole character down to zero energy (except for a few discrete points), as shown in Figs. \ref{FIG1} and \ref{FIG2}. This property can have measurable consequences in a tunneling experiment. It has been shown that, in the presence of dissipation (e.g., from a parent superconductor with sub-gap states),\cite{Bauriedl1981,Yazdani1997,Martin2014,Liu2017,Stenger2017} the differential conductance is particle-hole asymmetric. Essentially, a state of energy $E_n$ with, say, particle character and its hole-type counterpart at energy $-E_n$ will generate differential conductance signals of different amplitudes. This asymmetry does not emerge in the case of a split Majorana mode (as long as the splitting is not too large), because  both the positive- and negative-energy states are equal mixtures of particle and hole components. By contrast the signature of an ABS generated by the inter-band coupling mechanism can be particle-hole asymmetric (in the presence of dissipation) down to arbitrarily low energy. To illustrate this point, in Fig. \ref{FIGx} we compare the differential conductance trace  generated (in the presence of finite dissipation)
by an ABS pinned near zero energy by level repulsion [panel (a)] and the trace associated with an oscillating Majorana mode [panel (b)]. Note the manifest low-energy particle-hole asymmetry in panel (a). More specifically, the asymmetric zero bias peak extending from $\Gamma \approx 0.7~$meV to $\Gamma \approx 1.0~$meV in panel (a) is a clear sign that the underlying BdG state has asymmetric particle-hole character down to very small energies. We emphasize that the asymmetric features discussed here are characterized by an energy scale on the order of the ZBCP width. Asymmetric features occurring at higher energies (inside or outside the induced gap), do not provide any information regarding the Majorana (or non-Majorana) nature of the lowest energy mode.       
Based on the fundamental property discussed here, we conclude that any \textit{nearly}-zero-bias differential conductance feature that does not exhibit particle-hole symmetry should not be attributed to MZMs, quasi-Majoranas, or any other low-energy mode that involves (partially) separated Majorana bound states, but rather to the presence of (topologically-trivial) ABSs pinned near zero-energy by level repulsion.  

\subsection{Inhomogeneous wires} \label{Inhomo}
A position-dependent effective potential causes variations (along the wire) of the transverse profiles of the wave functions associated with different confinement-induced bands, which, in turn, induces inter-band coupling.\cite{Woods2018}
Inhomogeneous potentials can arise in Majorana devices for reasons such as the termination of the SC covering, the use of multiple electrostatic gates along the wire, an inhomogeneous SC-SM work function difference, the presence of multiple leads used in tunneling spectroscopy, and disorder. We note that the presence of an inhomogeneous potential can induce trivial low-energy ABSs that stick to zero energy even within single-band models (i.e. ps-ABSs generated via the partial-separation mechanism), as discussed extensively in the literature.\cite{Reeg2018b,Moore2018a,Stanescu2018b}
Here, we show that the inter-band-coupling induced by an inhomogeneous potential can also lead to the emergence of ABSs pinned near zero energy, but these ABSs are composed of non-separated Majorana components. 
Note that the  inhomogeneous potential also breaks the inversion symmetry of the system, which allows the presence of an anti-symmetric component of  the longitudinal spin orbit coupling in the toy model, i.e. $\widetilde{\alpha}^{\alpha\beta}_i =  -\widetilde{\alpha}^{\beta\alpha}_i$ for $\alpha \neq \beta$ (see Appendix \ref{App1}). As a consequence, transverse spin-orbit coupling is not needed to induce inter-band level repulsion within inhomogeneous systems, in contrast to the homogeneous case.\cite{Moor2018}

\begin{figure}[t]
\begin{center}
\includegraphics[width=0.48\textwidth]{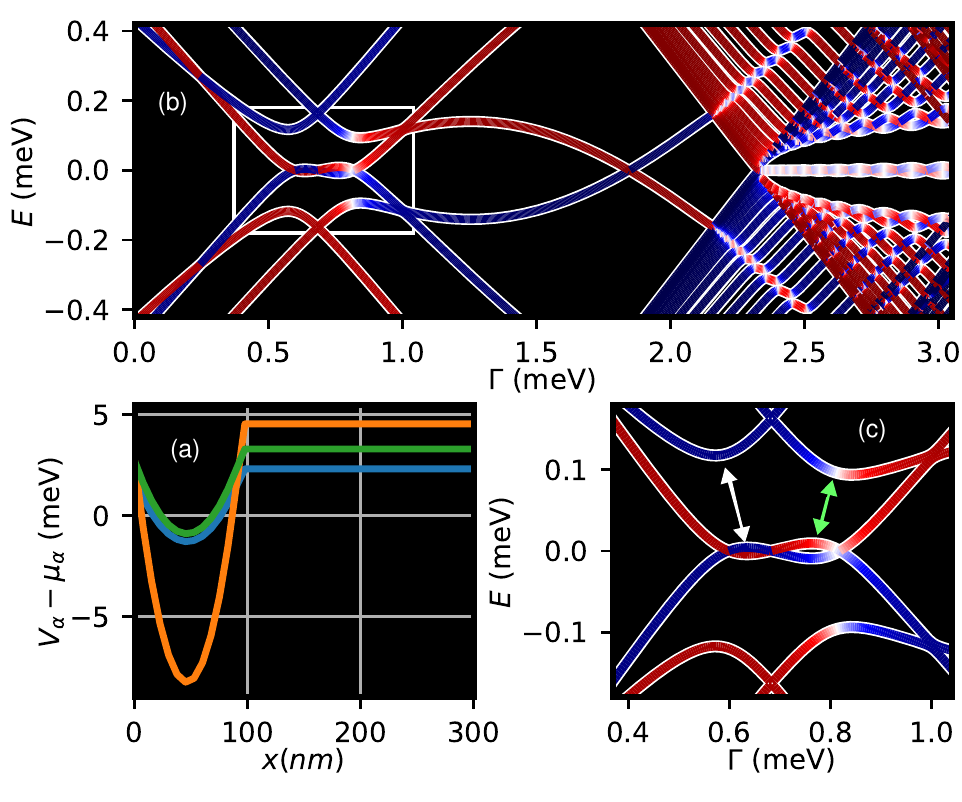}
\end{center}
\vspace{-5mm}
\caption{(a) Band-dependent effective potential (shifted by $\mu_\alpha$) as a function of position for a wire of length $L = 1.5~\mu$m in the presence of a gate potential at the left end of the system. We have $\mu_1 =-2.31~$meV (blue), $\mu_2=-3.3~$meV (green), and $\mu_3 =-4.54~$meV (orange). Within the gate region ($0\leq x\leq 100~$nm), $V_{i\alpha}$ is assumed to be harmonic with maximum depths of $3.6~$meV, $4.2~$meV, and $12.8~$meV, respectively, while the effective potential is zero outside of the gated region. Note that effective potential is shown only for the leftmost $300~$nm. (b) Dependence of the low-energy spectrum on the applied Zeeman field. Panel (c) represents a  zoomed in look of the region outlined by a white box in panel (b).
The model parameters are: $\alpha_{1,1},\alpha_{2,2},\alpha_{3,3} = 500, 333,$ and $250$ meV \AA, respectively, $\alpha_{1,2} = -\alpha_{2,1} = 5$ meV \AA, $\Delta_{i,i} = 0.35$ meV, and $\Delta_{1,3} = \Delta_{3,1}= 0.175$ meV. }
\label{FIG4}
\vspace{-3mm}
\end{figure}

Consider a wire of total length $L = 1.5$ $\mu$m described by a three band model. We assume all three bands to be empty in the bulk of the wire at zero magnetic field. We also assume a potential well localized near the left end of the wire, within a $100~$nm long region, where the three bands become occupied (i.e. $V_\alpha < \mu_\alpha$) as the result of applying a gate potential, as shown in Fig. \ref{FIG4}(a).
We include a key insight from the 3D model by making the effective potential band-dependent within the gated region (see Sec. \ref{3Dmodel}). This is due to the fact that that the three bands have, in general, different transverse profiles.
For example, the band corresponding to the orange curve
in Fig. \ref{FIG4}(a) may have more weight near the gate, as compared to the other two bands, and, therefore, its effective potential is more affected by the applied gate potential. The inter-band coupling includes terms associated with the (induced) superconducting pairing and the longitudinal spin-orbit coupling, as predicted by the 3D model.

The dependence of the  low-energy spectrum on the applied Zeeman field is shown in Fig. \ref{FIG4}(b).  The first notable feature is represented by the sub-gap states generated by the inhomogeneous potential at low fields, i.e. in the topologically trivial phase characterized by $\Gamma < \Gamma_c\approx 2.3~$meV. The bulk quasiparticle gap vanishes at the TQPT corresponding to $\Gamma=\Gamma_c$, then it reopens,  simultaneously with the emergence of MZMs. 
While the existence of ABSs induced by short-range inhomogeneous potentials (which can cross zero energy in the topologically-trivial regime\cite{Moore2018}) was discussed in the literature, a remarkable feature of this ABS mode is the pinning near zero energy over a considerable Zeeman field range [$\sim 0.3~$meV; see Fig. \ref{FIG4}(c)], despite the very short length scale associated with the inhomogeneity (about $100~$nm). We emphasize that a single band model with similar parameters predicts zero-energy ABS crossings, but no pinning over a finite Zeeman field range.\cite{Moore2018} In fact, the low-energy ABS illustrated in Fig. \ref{FIG4},  which sticks near zero energy over a substantial range of Zeeman field, is generated by the  inter-band coupling mechanism discussed above, hence it requires multi-band occupancy. Reproducing this behavior within a single band model would require an inhomogeneity with characteristic length scale on the order of a micron.\cite{Moore2018}

Analyzing the structure of the low-energy states associated with the spectrum shown in Fig. \ref{FIG4} provides us with a physical picture of the inter-band coupling mechanism responsible for the pinning of the low-energy mode. For example, in the vicinity of the first zero energy crossing near $\Gamma=0.6$ meV, the lowest energy state has most of its weight coming from the band shown in green in Fig. \ref{FIG4}(a). In the absence of inter-band coupling, this state simply crosses zero energy and leaves the energy window represented in Fig. \ref{FIG4}(b) near $\Gamma=1$ meV. However, in the presence of inter-band coupling, the state hybridizes with another low energy state associated (primarily) with the band  corresponding to the `blue' effective potential in panel (a), which results in the anti-crossing  indicated in Fig. \ref{FIG4}(c) by the white arrow. Note that both of these states are hole-like (blue filling) -- here, as before, we focus on the positive energy states -- and maintain their hole character throughout the anti-crossing. The primary mechanism responsible for this anti-crossing is the inter-band spin-orbit coupling $\alpha_{1,2}$ between the `blue' and `green' bands, which are the main components of the two states. By contrast, the second anti-crossing indicated by the green arrow in Fig. \ref{FIG4}(c) corresponds to a particle-like and a hole-like state (associated, primarily, with the `green' and `orange' bands, respectively) approaching each other. The particle-hole characters are exchanged between the two states through the anti-crossing. Since particle-hole coupling occurs due to SC pairing, we conclude that the primary mixing mechanism responsible for this anti-crossing is the inter-band superconducting pairing  $\Delta_{13}$ between the `green' and the `orange' bands of Fig. \ref{FIG4}(a).

\begin{figure}[t]
\begin{center}
\includegraphics[width=0.48\textwidth]{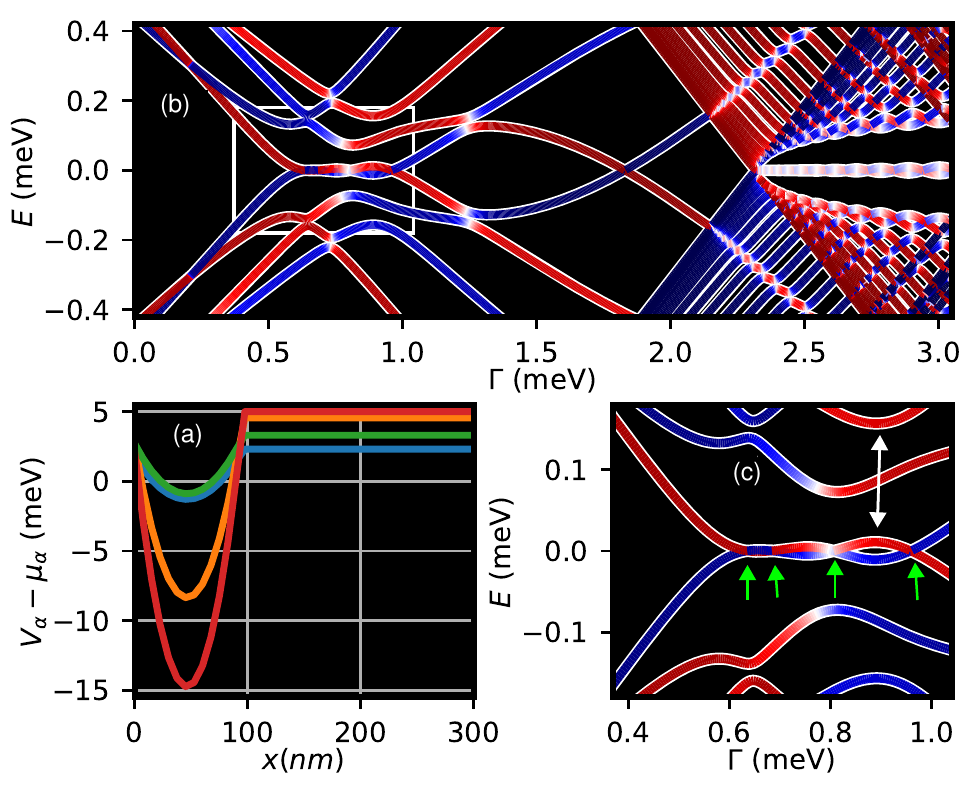}
\end{center}
\vspace{-5mm}
\caption{(a) Band-dependent effective potential (shifted by $\mu_\alpha$) as a function of position for a wire of length $L = 1.5~\mu$m in the presence of a gate potential at the left end of the system. The parameters corresponding to the first three bands are the same as in Fig. \ref{FIG4}. The forth band (red) is characterized by $\mu_4 =-5~$meV, while $V_{i,4}$ is a harmonic well inside the gate region with a maximum depth of $19.75~$meV and is zero outside. The other parameters associated with the fourth band are: $\alpha_{4,4} = 333$ meV \AA, $\alpha_{1,4} =-\alpha_{4,1}= 5$ meV \AA, $\alpha_{3,4} = -\alpha_{4,3}=72.5$ meV \AA, $\alpha_{2,4}=-\alpha_{4,2}=10$ meV \AA, and $\Delta_{3,4} = \Delta_{4,3}=0.1$ meV. (b) Dependence of the low-energy spectrum on the applied Zeeman field.  Panel (c) represents a  zoomed in look of the region outlined by a white box in panel (b).}
\label{FIG5}
\vspace{-3mm}
\end{figure}

A more robust (nearly) zero-energy state can be obtained by adding a fourth band to the model described above,  more specifically the `red' band in Fig. \ref{FIG5}(a), which is  characterized by a deep potential well in the barrier region. The corresponding low-energy spectrum is shown in Fig. \ref{FIG5}(b),  with panel (c) representing a zoomed in view of the low-field near-zero ABS. Again, the lowest energy mode is pinned near zero energy over a significant range of Zeeman field ($\sim 0.4~$meV) in the topologically trivial regime. Note that experimentally the high-field regime (e.g., $\Gamma > 1.5~$meV)  may be inaccessible due to the collapse of the SC gap of the parent superconductor, so that the most prominent  low-energy feature would be the ZBCP generated by the topologically-trivial ABS  mode pinned near zero energy.
The expanded pinning range (as compared with the three-band model shown in  Fig. \ref{FIG4}) is due to the additional anti-crossing marked by the white arrow in Fig. \ref{FIG5}(c), which is primarily due to inter-band spin-orbit coupling (described primarily by $\alpha_{3,4}$), since the relevant states have particle-like character through the entire anti-crossing (i.e. red filling at positive energies -- see Fig. \ref{FIG5}). 

\begin{figure}[t]
\begin{center}
\includegraphics[width=0.48\textwidth]{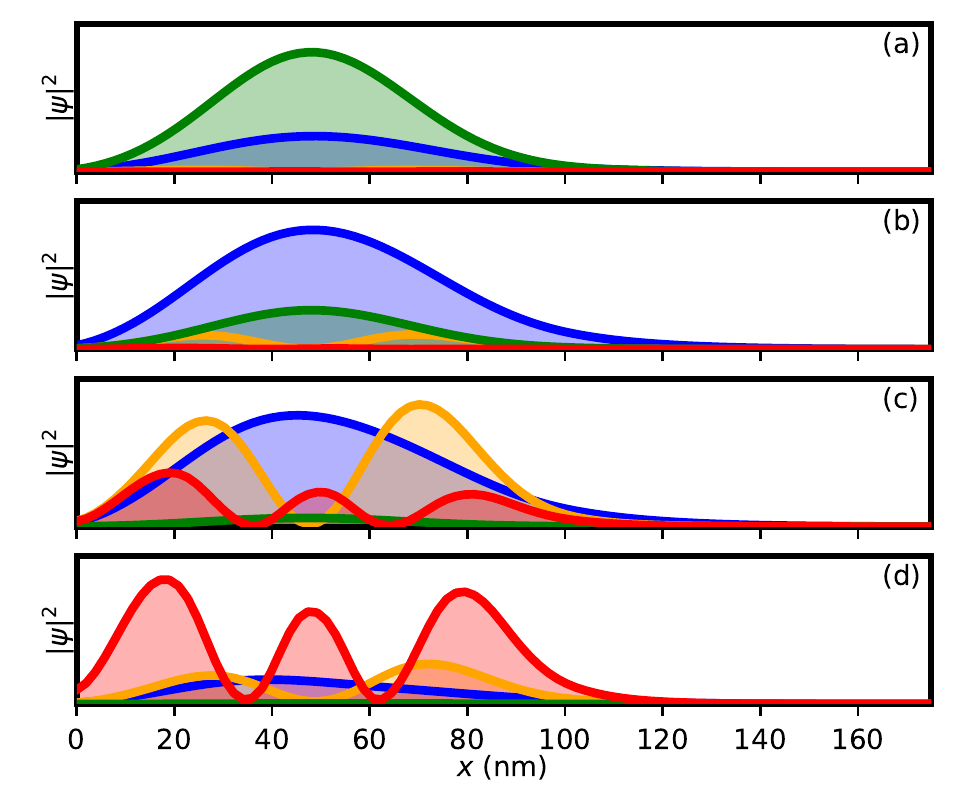}
\end{center}
\vspace{-5mm}
\caption{Wave function profiles corresponding to the zero-energy states marked by green arrows in Fig. \ref{FIG6}(c). The calculated amplitudes $|\psi|^2$  involve summations over the spin and particle-hole degrees of freedom. Different colors represent contributions from the corresponding bands, using the color code from Fig. \ref{FIG5}(b). Note that the low-energy ABS mode is localized within the gate region ($0\leq x\leq 100~$nm) and that its band composition changes dramatically with the Zeeman field as a result of strong inter-band coupling.}
\label{FIG6}
\vspace{-3mm}
\end{figure}

To demonstrate that the anti-crossings are indeed due to inter-band coupling, we calculate explicitly the wave function amplitudes corresponding to the four zero energy crossings marked by the green arrows in Fig. \ref{FIG5}(c). The results are shown in Fig. \ref{FIG6}. Note that the contributions to the lowest-energy mode from various confinement-induced bands change quite dramatically as the Zeeman field increases, which is a clear indication of inter-band mixing. In addition, the explicit calculation of the corresponding Majorana components (see Appendix \ref{App2}) reveals the absence of any significant Majorana separation, which confirms that the partial separation mechanism is not responsible for the pinning of this ABS mode near zero energy. 

\begin{figure}[t]
\begin{center}
\includegraphics[width=0.48\textwidth]{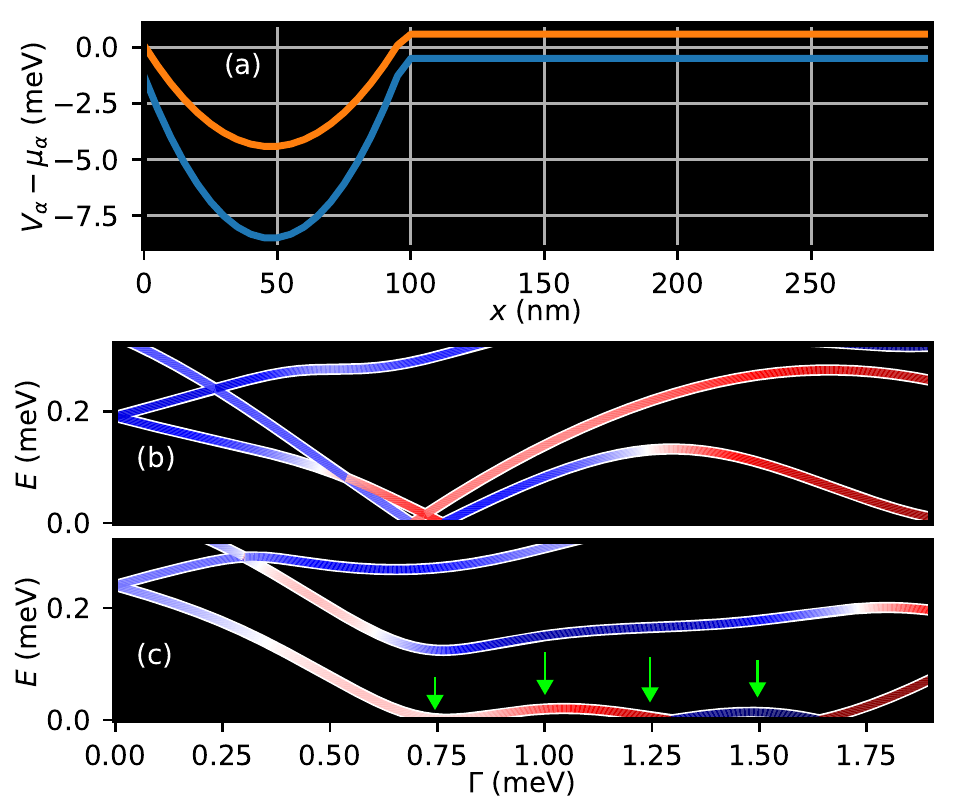}
\end{center}
\vspace{-5mm}
\caption{The system is described by a two-band model with a position- and  band-dependent effective potential (shifted by $\mu_\alpha$) shown in panel (a).  The chemical potential is place between the two bands, such that $\mu_1=0.5~$meV and $\mu_2=-0.57~$meV. The potential well depths are $8~$meV and $5~$meV, respectively. Dependence of the low-energy spectrum on the applied Zeeman field for a wire of length $L = 0.3~\mu$m in the presence of a gate potential at the left end of the wire for (b)  a system without inter-band coupling and (c) a system with inter-band spin-orbit coupling. The other model parameters are: $\Delta_{11}=\Delta_{22}=0.35~$meV (only within the proximitized, homogeneous potential region) and, for panel (c), the inter-band coupling is given by ($\alpha_T)_{12}=1.2~$meV (only within the inhomogeneous, gate region).}
\label{FIG7}
\vspace{-3mm}
\end{figure}

\begin{figure}[t]
\begin{center}
\includegraphics[width=0.48\textwidth]{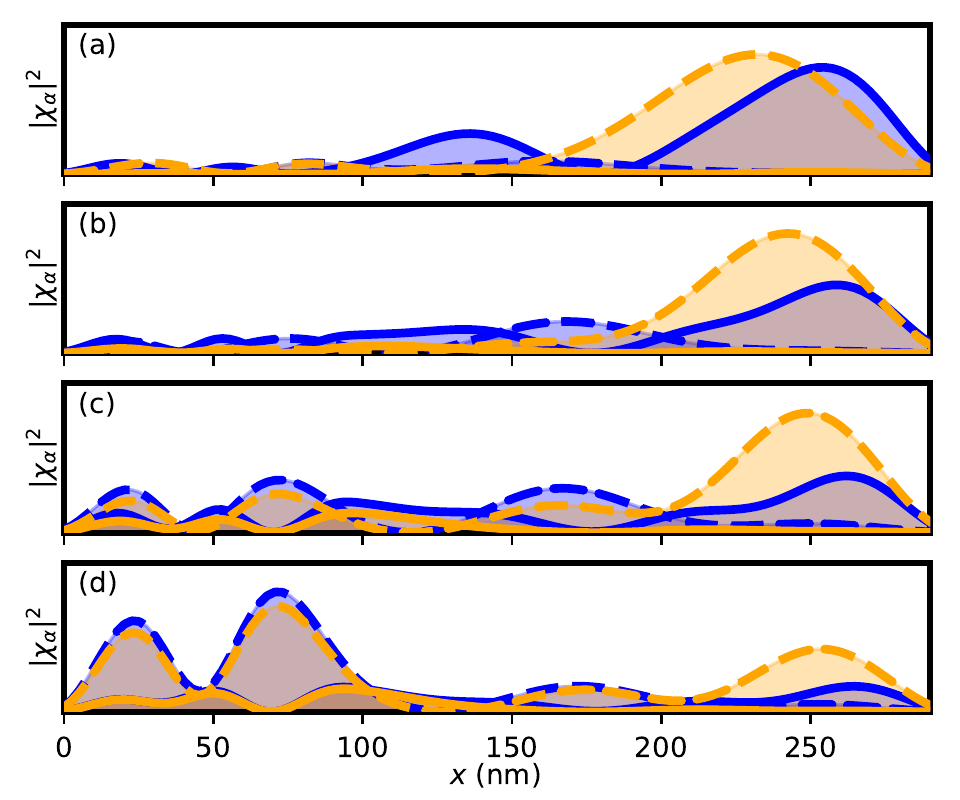}
\end{center}
\vspace{-5mm}
\caption{Majorana wave function profiles corresponding to the low-energy states marked by green arrows in Fig. \ref{FIG7}(c). The two Majorana wave functions are shown in blue and orange, respectively, while solid and dashed lines indicate the contributions from the first and second bands, respectively.}
\label{FIG8}
\vspace{-3mm}
\end{figure}

As a final example, we  consider a two-band model representing a short wire of length $L=300~$nm with an inhomogeneous potential as shown in Fig. \ref{FIG7}(a). This model differs from the previous two examples in two respects: (1) the inter-band coupling is active only within the inhomogeneous potential (gate) region $0\leq x \leq 100~$nm and (2) the induced superconductivity is nonzero only outside of the gate region. The model corresponds to a setup consisting of a short proximitized wire coupled to a quantum dot representing the uncovered gate region. 
As the gate-induced potential in the dot region is highly  inhomogeneous, we expect the emergence of strong inter-band mixing, as confirmed by the 3D calculation (see Sec. \ref{3Dmodel}). The dependence of the low-energy spectrum on the Zeeman field is shown in Fig. \ref{FIG7} for a system with decoupled bands, i.e., without inter-band coupling [panel (b)], and a system with inter-band spin-orbit coupling [panel (c)]. Note that in the proximitized region both band minima are relatively close to the chemical potential, $\left|\mu_i\right|\approx0.5~$meV, so that in the absence of inter-band coupling they satisfy the topological condition at about the same critical field.
However, due to the very short length of the wire, the two pairs of MBSs (one for each band) overlap strongly and the resulting energy splittings have amplitudes comparable to the induced gap, as shown in Fig. \ref{FIG7}(b). By contrast, inter-band spin-orbit coupling within the uncovered (gate) region  pins the lowest energy mode near zero energy over a very wide range of Zeeman energy from $\Gamma \approx 0.75~$meV to $\Gamma \approx 1.6~$meV. The nature of this low energy mode is revealed by calculating the wave functions of its Majorana components.
The amplitudes of the Majorana wave functions corresponding to the  low-energy states marked by green arrows in Fig. \ref{FIG7}(c) are shown in Fig. \ref{FIG8}. In panel (a), which corresponds to $\Gamma=0.75~$meV,  we notice two Majorana modes (orange and blue, respectively) localized near the right end of the wire. While they overlap strongly, these modes belong to different bands, as indicated  by the solid (first band) and dashed (second band) lines. This result can be understood as follows: in the absence of inter-band coupling, the system supports two pairs of (strongly overlapping) MBSs associated with the two bands.  When the inter-band coupling is turned on in the inhomogeneous region, the two Majoranas localized near the left end of the system get coupled and morph into a finite energy ABS. This leaves two unpaired Majoranas at the right end of wire that are spatially overlapping, but are (partially) separated in band space. 
Note that, in the short wire studied here, the two Majoranas still have tails leaking into the region where the bands couple. Naively, one would expect significant energy splitting oscillations as a result of this coupling. However, upon increasing $\Gamma$, the lowest energy state is transforming as a result of inter-band coupling from a pair of band-separated Majoranas localized near the right end of the wire [panel (a)] into a low-energy ABS (predominantly) localized within the uncovered region and characterized by strongly overlapping Majorana modes associated with the second band [panel (d)]. Hence, the robust pinning near zero energy shown in Fig. \ref{FIG7}(c) is due to a combination of two distinct mechanisms: the band separation of the MBSs  localized in the homogeneous (uncoupled) region and the level repulsion (induced by inter-band coupling) affecting the ABS localized within the uncovered region. We will discuss a similar example within the 3D model, at the end of Sec. \ref{3D Results}. The important message here is that, in general, the inter-band coupling mechanism acts in combination with the partial separation (in real space or band space) mechanism. A combination of these mechanisms in very short systems can result  in a rather spectacular pinning of the lowest-energy mode near zero energy.  This example further demonstrates that the observation of low-amplitude energy splitting oscillations is not necessarily an indication of topological protection.

\section{3D model} \label{3Dmodel}
The 3D model calculations incorporate electrostatic effects due to gate-induced external potentials and the presence of a parent superconductor. Their purpose is twofold. First, we want to understand if the basic assumptions underlying the simplified models used so far in the literature for discussing topologically-trivial ABSs emerging in hybrid systems, 
or those underlying the inter-band coupling mechanism discussed above
are realistic enough. For example, does a specific experimental setup generate an electrostatic confinement that is smooth enough to induce robust ps-ABSs?
More importantly for this work, are the actual inter-band couplings strong enough to trigger the level repulsion mechanism discussed in the previous section? Our second purpose
is to estimate whether the emergence of low-energy ABSs due to level repulsion is a rather generic occurrence, or rather one that requires a lot of fine tuning. 

\subsection{Theoretical Model} \label{3D theory}
\begin{figure}[t]
\begin{center}
\includegraphics[width=0.48\textwidth]{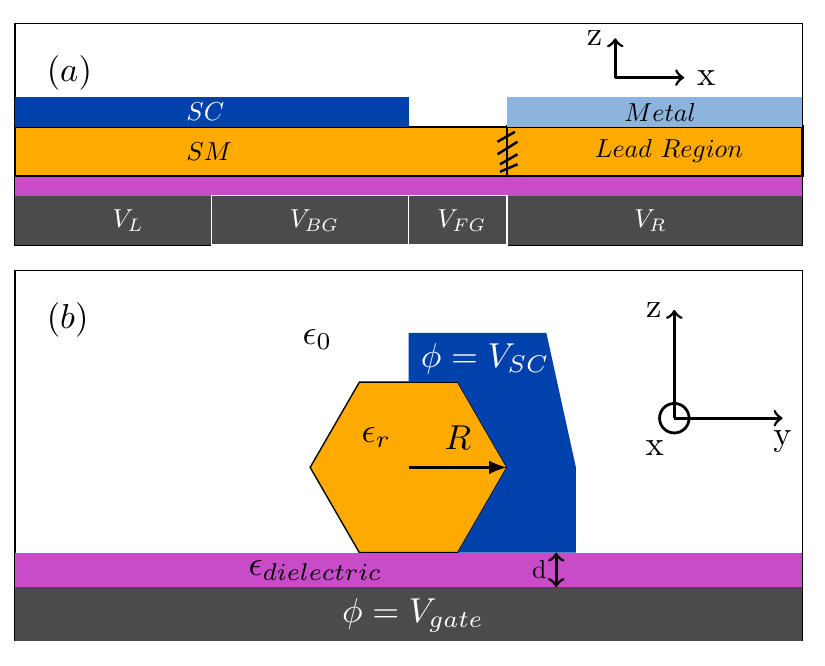}
\end{center}
\vspace{-5mm}
\caption{(a) Schematic representation of the device along the SM wire. The semiconductor nanowire (orange) is proximity coupled to a superconductor (blue) and a metal (light blue) is the lead region. A dielectric layer (purple) separates the nanowire from potential gates (dark grey). There are four regions defined by the external gates and the materials deposited on the SM wire: the left bulk region  (gate potential $V_L$),  the big-gate region ($V_{BG}$), the fine-gate (uncovered) region ($V_{FG})$, and the lead region ($V_R$). The lead region is a continuation of the SM nanowire but in proximity to a metal, which strongly renormalizes its properties. We focus on the physics of the isolated hybrid nanowire, which is disconnected  from the lead (as indicated by black lines between the SM and the lead region). (b) Schematic representation of the cross section of the nanowire device in the SC-covered region.}
\label{FIG10}
\vspace{-3mm}
\end{figure}

In this section we describe the 3D model used to study the effects of an inhomogeneous electrostatic potential and the multi-band physics in Majorana nanowires. A schematic representation of a setup that matches the devices used in a recent tunneling experiment \cite{Chen2019} is shown in Fig. \ref{FIG10}. The basic ingredients include a semiconductor wire (SM) in proximity to an s-wave superconductor (SC), a normal lead (used for tunneling spectroscopy), and various gates to control the electrostatic potential, as shown in panel (a). The transverse profile of the system in the SC covered region in shown in panel (b). Note that the SC is treated as a boundary condition, as far as  the electrostatic effects are concerned,  with a potential $V_{SC}$ set by the work function difference between the SC and SM. The details of the lead region (which consists of a SM wire segment covered by normal metal) are quite difficult to model due to the unknown parameters characterizing this region. The metallic material alters the electrostatic conditions (due to the work function difference between the metal and the SM) and heavily renormalizes the effective parameters of the wire due to strong hybridization between SM and metallic states. Accurately capturing these effects would require to explicitly incorporate the normal metal into the model. 
Since we are not concerned here with tunneling features,  we focus on the physics of an isolated nanowire (i.e. a nanowire that is not coupled to a tunneling probe). However, we still take into account electrostatic effects due to the presence of the lead. 
The lead region is disconnected from the SM region (as indicated by black lines), but the metal covering (light blue region above lead region) is still incorporated when calculating the external electrostatic potential. This allows us to include the electrostatic screening effect of the lead, without explicitly incorporating the normal metal into the Hamiltonian.

There are four different gates with potentials denoted by $V_L$, $V_{BG}$, $V_{FG}$, and $V_R$ respectively. These gates break up the device into four regions: the left bulk region  (gate potential $V_L$),  the big-gate region ($V_{BG}$), the fine-gate (uncovered) region ($V_{FG})$, and the lead region ($V_R$).  Our focus is on the low-energy physics of the BG and FG regions, which can be probed using tunneling spectroscopy (from the right lead). Consequently, the gate potential $V_L$ is set such that the low energy states do not leak significantly into the left bulk region. 
The low-energy states of interest are therefore confined to the BG and FG regions with the corresponding gates being used as control knobs for the electrostatic potential within these active regions. 

The tight-binding model used in the 3D electrostatic calculations is constructed\cite{Woods2018} by dividing the semiconductor into $N_x$ layers along the length of the wire, each containing $N_\bot$ sites. 
The corresponding Hamiltonian, which does not include superconductivity, is given by
\begin{equation}
\begin{aligned}
H_{SM} =& \sum\limits_{i,j,m,\sigma} t^{\bot}_{ij} c_{im\sigma}^{\dagger}c_{jm\sigma}
 + \sum\limits_{i,m,n,\sigma} t^{\parallel}_{mn} c_{im\sigma}^{\dagger}c_{in\sigma} \\
 & + \sum\limits_{i,m,\sigma} \left(V_{im}+U_{im}\right) n_{im\sigma} \\
  & + \sum\limits_{i,m,\sigma,\sigma^{\prime}} \alpha_{R}  \left[c_{i(m+1)\sigma}^{\dagger} \left(i\sigma_{y}\right)_{\sigma\sigma^{\prime}} c_{im\sigma^{\prime}}  + h.c. \right]\\
  & +\sum\limits_{i,m,\sigma,\sigma^{\prime}} \Gamma~ c_{im\sigma}^{\dagger} \left(\sigma_{x}\right)_{\sigma\sigma^{\prime}} c_{im\sigma^{\prime}}, \label{H3D}  
\end{aligned}
\end{equation}
where $c_{im\sigma}^{\dagger}$ creates an electron with spin $\sigma$ localized near the site $i$ of layer $m$,  $n_{im\sigma} = c_{im\sigma}^{\dagger}c_{im\sigma}$ is the number operator,
$t^{\bot}_{ij}$ and $t^{\parallel}_{mn}$ are intra- and inter-layer nearest neighbor hopping matrix elements, respectively, $\Gamma$ is the (half) Zeeman splitting, and  $\alpha_{R}$ is the Rashba spin-orbit coefficient. Note that we neglect transverse spin-orbit coupling, for simplicity.
The electrostatic effects due to the presence of the potential gates and  the superconductor are described by the external potential $V_{im}$. Explicitly, the potential matrix elements are $V_{im}= -e\left<i,m\left|V(\bold{r})\right|i,m\right>$, where $V(\bold{r})$ is the solution of the  Laplace equation $\nabla^2V(\bold{r}) = 0$ with boundary conditions set by the potential on the superconductor ($V_{SC}$) and the external gates ($V_{BG}$, $V_{FG}$, etc.). Electron-electron interactions are included at the mean field level through the term $U_{im}$.  Explicitly, the potential matrix elements are $U_{im}= -e\left<i,m\left|U(\bold{r})\right|i,m\right>$, where $U(\bold{r})$ is the solution of the Poisson equation $-\nabla^2\left[ \epsilon(\bold{r})U(\bold{r})\right] = \rho(\bold{r})$ with homogeneous boundary conditions.

The Schr\"{o}dinger-Poisson problem defined by the 3D Hamiltonian in Eq. (\ref{H3D} ), which necessarily involves a large number of degrees of freedom, can be efficiently solved by reducing it to an effective 1D problem though a projection onto an appropriate low-energy sub-space.  Here, we give a brief outline of the projection technique; the details of this procedure can be found in Ref. \onlinecite{Woods2018}. The essential observation behind this low-energy projection approach is that the transverse profiles of the (low-energy) states of a finite wire are quite similar to those of an infinite homogeneous wire with electrostatic environment similar to the local environment of the finite system. To incorporate this observation, we define an auxiliary Hamiltonian for each layer:
\begin{equation}
\begin{gathered}
H^{(m)}_{aux} = \sum\limits_{i,j,k,\sigma} \!\!\left[ t^{\bot}_{ij} + \left({\hbar^{2}k^{2} \over 2m^{*}} 
\!+\!V_{i}^{(m)} \!\!+\! U_{i}^{(m)} \right) \delta_{ij} \right] c_{ik\sigma}^{\dagger} c_{jk\sigma} \\
 +\sum\limits_{ik\sigma\sigma^{\prime}} \alpha_{R}k ~c_{ik\sigma}^{\dagger} \left(\sigma_{y}\right)_{\sigma\sigma^{\prime}} c_{ik\sigma^{\prime}}, \label {Haux}
 \end{gathered}
\end{equation} 
where $V_i^{(m)}=V_{im}$. The auxiliary Hamiltonian of the $m^{th}$ layers describes an infinite wire with a translation-invariant external potential that matches the local external potential of the finite wire and whose transverse profile also matches the local transverse profile of the $m^{th}$ layer. 
The low-energy  $k=0$ eigenstates of the effective Hamiltonian can be viewed as a set of `molecular orbitals' and provide us with a  position-dependent (i.e. layer-dependent) basis for the low-energy sub-space. Finally, the low-energy effective 1D Hamiltonian is obtained by projecting the full 3D Hamiltonian, Eq. (\ref{H3D}), onto the sub-space spanned by the $n_o$ lowest molecular orbitals [i.e. $k=0$ eigenstates of Eq. (\ref{Haux})]. Explicitly, we have
\begin{equation}
\begin{gathered}
H_{SM}^{\rm eff} = \sum_{m,n,\sigma}\sum_{\alpha,\beta} ^\bullet \tilde{t}^{\parallel}_{m\alpha,n\beta} ~ {c}_{m\alpha\sigma}^{\dagger}{c}_{n\beta\sigma} +\sum_{m,\sigma}\sum_{\alpha} ^\bullet \epsilon_{\alpha}^m ~{n}_{m\alpha\sigma} \\
+\sum_{m,\sigma\sigma^\prime}\sum_{\alpha,\beta} ^\bullet \left[\widetilde{\Delta U}_{\alpha \beta}^{m}~\delta_{\sigma\sigma^\prime} + {\Gamma}\left(\sigma_{x}\right)_{\sigma\sigma^{\prime}}\delta_{\alpha\beta}\right]c_{m\alpha\sigma}^{\dagger}  c_{m \beta \sigma^{\prime}} \\
+\sum_{m,n,\sigma\sigma^\prime}\sum_{\alpha,\beta} ^\bullet i \alpha_{\alpha\beta}^{mn} (\sigma_y)_{\sigma\sigma^\prime} ~c_{m\alpha\sigma}^{\dagger}  c_{n \beta \sigma^{\prime}},     \label{Heff}
\end{gathered}
\end{equation}
where m and n label the sites of a 1D lattice, $\alpha$ and $\beta$ label the molecular orbitals corresponding to the eigenstates of $H_{aux}^{(m)}$, $\epsilon_\alpha^m$ are the energies of the molecular orbitals for layer m, and the summations marked by a $\bullet$ are restricted to the low energy subspace. 
The hopping matrix elements $\widetilde{t}_{m\alpha,n\beta}$ are given by
\begin{equation}
\tilde{t}^{\parallel}_{m\alpha,n\beta} = \langle\varphi_\alpha^m|T^\parallel|\varphi_\beta^n\rangle,
\end{equation}
where $\left[T^\parallel\right]_{im,in} = t_{mn}^{\parallel}\delta_{ij}$ and $|\varphi_\alpha^m\rangle$ , $|\varphi_\beta^n\rangle$ are eigenstate of $H_{aux}^{(m)}$ and $H_{aux}^{(n)}$, respectively. 
The spin-orbit matrix elements $\alpha_{mn}^{\alpha\beta}$ are calculated in a similar manner. Notice that the inter-band hopping and the inter-band spin-orbit coupling are nonzero if (and only if) the eigenstates  corresponding to  neighboring layers are different, i.e. if the transverse profile of the states associated with a given `molecular orbital' change as a function of position along the wire. This occurs when the effective potential is position-dependent. Consequently, we expect inter-band coupling to occur, for example, near the interface of the BG and FG regions [see  of Fig. \ref{FIG10}(b) and (c)] due to the termination of the SC covering and having (in general) different  values for $V_{BG}$ and $V_{FG}$. The quantity $\widetilde{\Delta U}_{\alpha \beta}^{m}$ describes the difference between the mean field potential of the auxiliary Hamiltonian $H_{aux}^{(m)}$ and the actual mean field potential of the 3D Hamiltonian given by Eq. (\ref{H3D}). This term arises primarily due to charge redistribution along the length of the wire and can lead to barrier-like features between the SC-covered and the uncovered regions.\cite{Woods2018} For simplicity, we neglect this term in the current analysis. In other words, we incorporate the mean field self-consistency when calculating the eigenstates of Eq. (\ref{Haux}), but neglect any fluctuation of the mean field value due to broken translation invariance or nonzero applied magnetic field. These fluctuations are important for quantitative considerations, however, the focus of this work is to illustrate the main qualitative features of multi-band physics in devices with inhomogeneous potentials. 

Lastly, we incorporate superconductivity at the mean field level through the pairing term
\begin{equation}
H_{\Delta}=\sum_{n,m,\alpha,\beta} \left[\langle\varphi_\alpha^n|\Delta|\varphi_\beta^m\rangle c_{n\alpha \uparrow}^\dagger c_{m\beta \downarrow}^\dagger + h.c.\right]
\end{equation}
with $\Delta_{ij}^{mn} = \Delta_i^m \delta_{m,n}\delta_{i,j}$, where $ \Delta_i^m$ is zero everywhere except at the SM-SC interface. Note that the FG region does not contribute to pairing due to the termination of the superconductor. Also note that, in general, the  inter-band  pairing $\Delta^{\alpha\beta}$  can become significant when the gate voltage is comparable to or larger than the superconductor-semiconductor work function difference.
The total effective BdG Hamiltonian becomes
\begin{equation}
H_{BdG} = H_{SM}^{\rm eff} + H_{\Delta}. \label{BdG}
\end{equation}
Note that the structure of the effective Hamiltonian is similar to the structure of the  toy model in Eq. (\ref{Htoy}). The major difference is that the parameters of the toy model (including the number of occupied bands, the profile of the effective potential, the inter-band coupling parameters, etc.) are largely arbitrary, while the parameters of the effective Hamiltonian are calculated based on the geometric and electrostatic properties of the device. The results presented below show that the conditions required by the mechanism described Sec. \ref{Toy} can be realized and are even likely to occur in experimental device.

\begin{figure}[t]
\begin{center}
\includegraphics[width=0.48\textwidth]{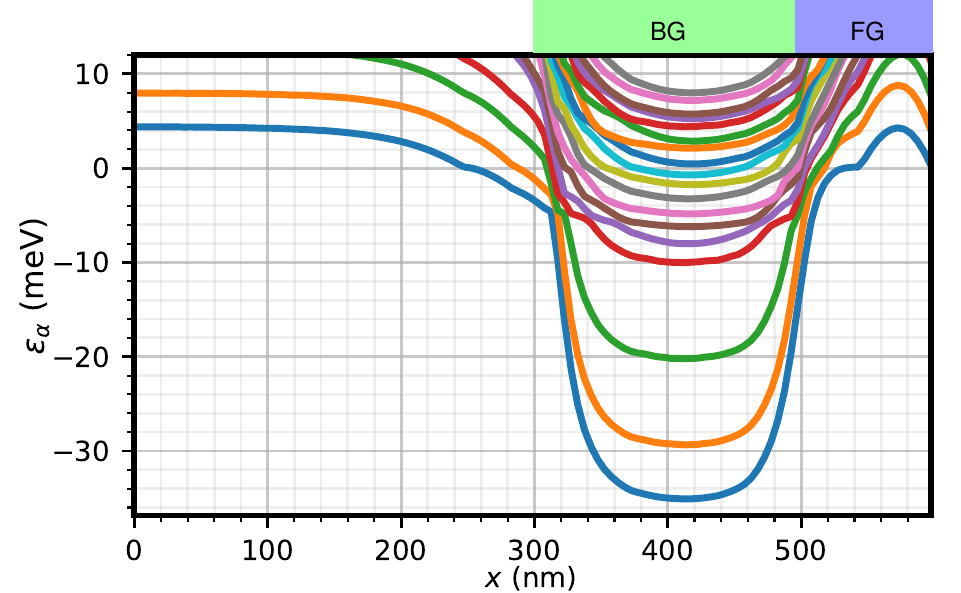}
\end{center}
\vspace{-5mm}
\caption{Effective potential profiles
corresponding to different bands for the device represented schematically in Fig. \ref{FIG10}. The BG region stretches from $300$ to $500~$nm, while the FG region stretches from $500$ to $600~$nm. The external  potentials are: $V_{SC}=230~$mV, $V_L=-250~$mV, $V_{BG}=364~$mV, $V_{FG}=175~$mV, and $V_R=-125~$mV.}
\label{3D_FIG2}
\vspace{-3mm}
\end{figure}

\subsection{Results} \label{3D Results}
The main findings discussed in this section are: i) for a wide range of experimentally-relevant gate potentials the system is characterized by multi-band occupancy (i.e. 5-20 occupied bands), ii)  within the 3D model, inter-band coupling arises naturally in the presence of inhomogeneities, and iii) inter-band coupling produces low-energy states that remain near zero energy over a wide range of Zeeman field due to inter-band level repulsion. Moreover, these ``sticky'' states occur quite frequently for systems with band occupancy of the order ten (and larger).

\begin{figure}[t]
\begin{center}
\includegraphics[width=0.48\textwidth]{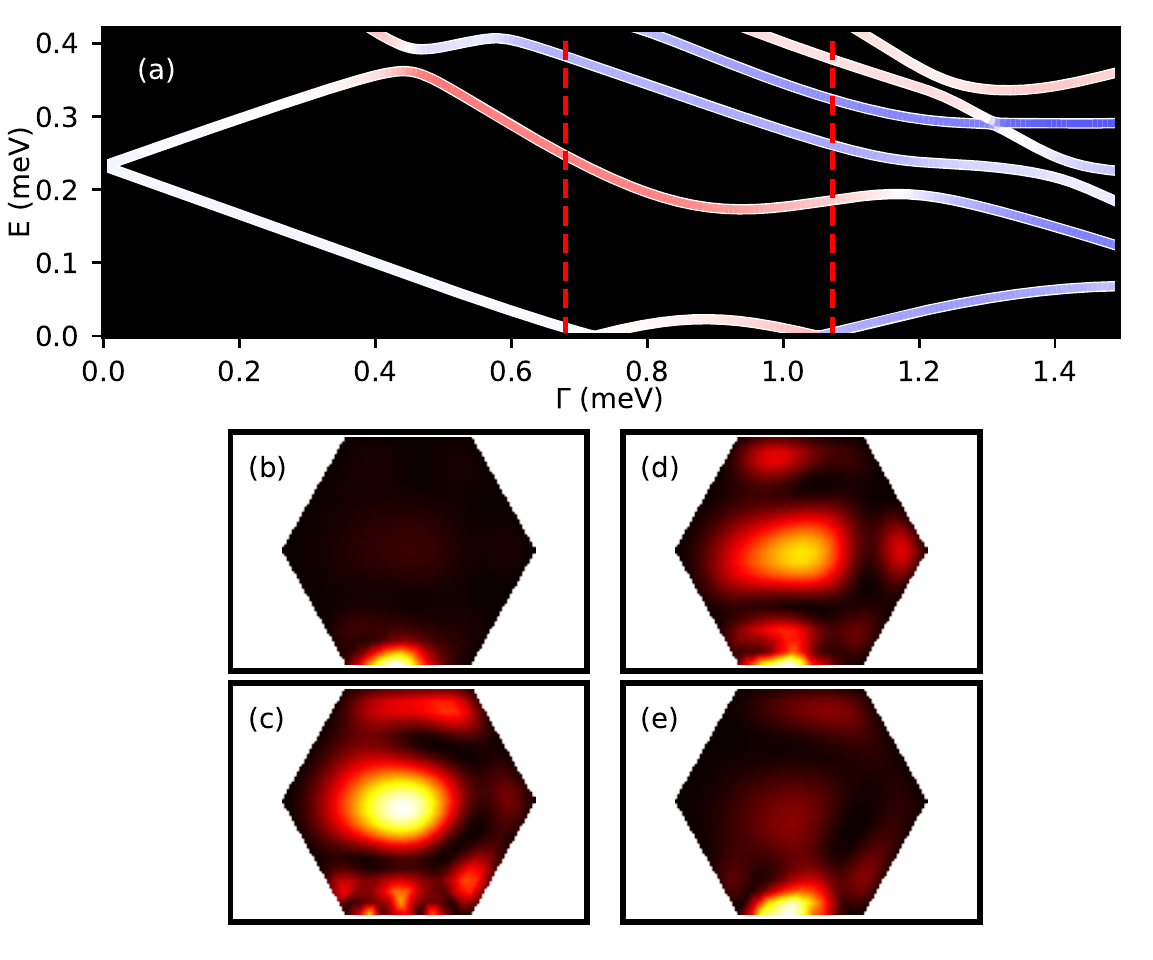}
\end{center}
\vspace{-5mm}
\caption{(a) Low-energy spectrum as a function of Zeeman field for a system with the same parameters as in Fig. \ref{3D_FIG2} (only the positive energy sector is shown). The transverse profiles of the first- and second-lowest energy states at $x=400~$nm (i.e. near the middle of the BG region, see Fig.\ref{3D_FIG2}) and Zeeman fields indicated by the red dashed lines in panel (a) are shown in panels (b)-(e). Panels (b) and (c) correspond to the states indicated by the first red dashed line, while (d) and (e) correspond to the second. Note that the states swap characters as the Zeeman field increases from $\Gamma\approx0.67~$meV to $\Gamma\approx1.13~$meV,  indicating anti-crossing behavior.}
\label{3D_FIG3}
\vspace{-3mm}
\end{figure}

Throughout this section, we use the following values for the system parameters: radius of the circle that circumscribes the SM wire $R = 70$ nm, 
thickness of the dielectric layer $d = 10~$nm, permittivity $\epsilon_r=17.7$ (wire) and $\epsilon_{dielectric} = 24$ (dielectric), effective mass $m^* = 0.025 m_{o}$, and Rashba coefficient $\alpha = 250$ meV \AA. The in-plane and out-of-plane lattice spacings are taken to be $a_\bot = 7$ nm and $a_\parallel = 5$ nm, respectively. The FG and BG regions are $100$ nm and $200$ nm long, respectively. The SC-SM work function difference is chosen to be $V_{SC} = 230$ mV and the gap between the bottom of the lowest-energy confinement-induced band and the chemical potential (before the external gates are applied) is $E_o=210~$meV. The geometric and dielectric parameters are chosen to match devices used in Ref.  [\onlinecite{Chen2019}]. 

The effective potential profile along the device corresponding to a given set of gate voltages is shown  in Fig. \ref{3D_FIG2}. Note that the BG and FG regions span from $300$ to $500~$nm and from $500$ to $600~$nm, respectively. As mentioned in the previous section, we are mainly interested in the low-energy physics of states localized (primarily) within the BG and FG regions. For this reason, $V_L$ has been set to a negative value, to suppress the leakage of low energy states in the left bulk region. There are several characteristics of the effective potential that deserve attention. Firstly, the spacing between successive bands is highly band-dependent within the BG region. In particular, the three lowest-energy bands  are widely separated (with inter-band gaps on the order of $10~$meV). These three bands have transverse profiles that are pinned near the BG gate, which explains why they  sink dramatically within the BG region upon applying a relatively strong (positive, i.e. attractive) gate potential. For the higher energy bands (fourth band and above) the inter-band spacing reduces to around $2~$meV or less, which dramatically increases inter-band coupling. 
These higher-energy bands are less confined near the BG gate, as compared to the lowest three bands. Secondly, 
one notices the rapid variation  of the effective potential near the edges of the BG region. This is caused, on the one hand, by the termination of the superconductor at the right edge of the BG region and, on the other hand, by the sudden change of the gate potential from $V_{BG}=364$ mV to $V_L=-250~$mV (at the left edge) and $V_{FG}= 175$ mV (at the right edge). This sharp variation of the effective potential causes several bands to cross zero energy and, very importantly, to switch order. This behavior, which is connected to a rapid evolution of the transverse profiles of the bands, is responsible for the large inter-band mixing that occurs within the transition regions.
 
\begin{figure}[t]
\begin{center}
\includegraphics[width=0.48\textwidth]{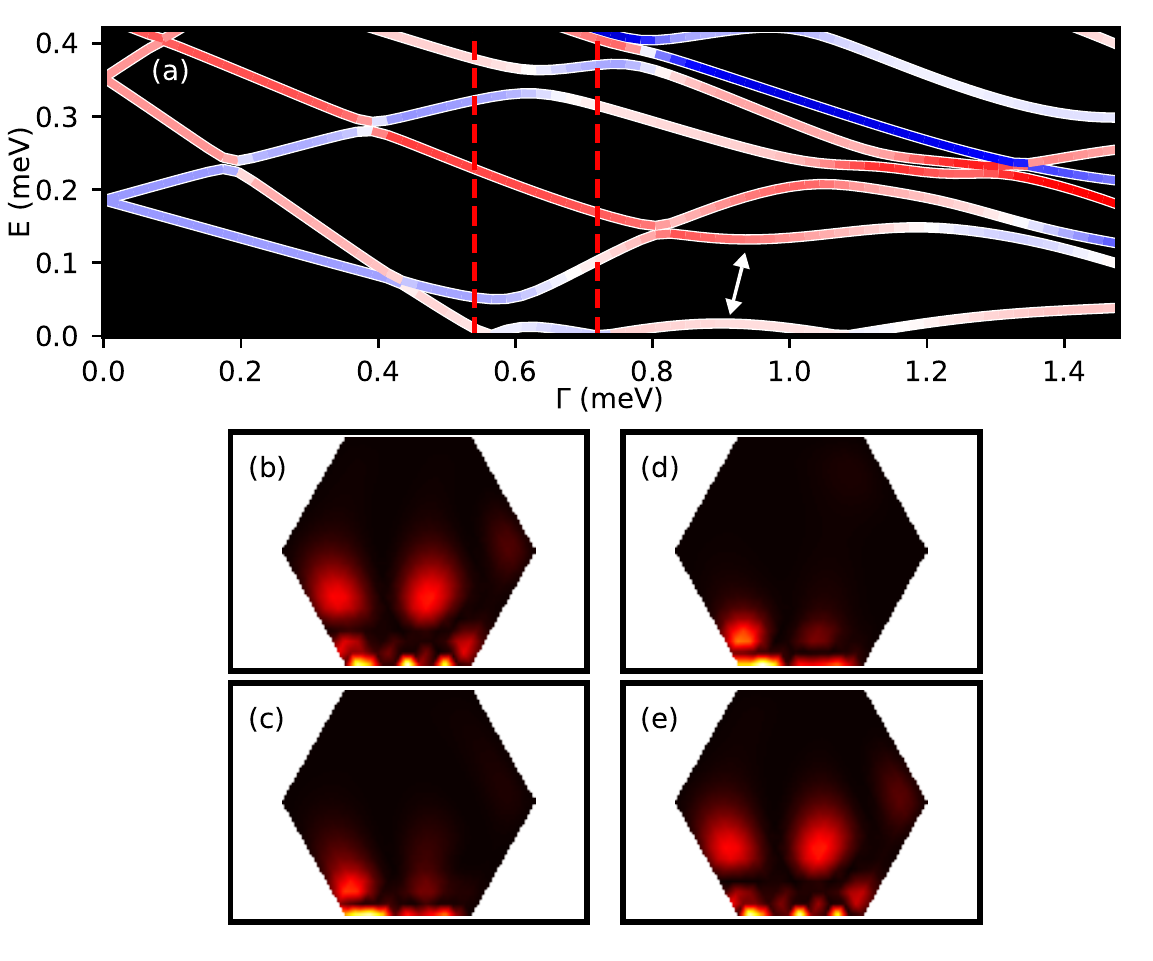}
\end{center}
\vspace{-5mm}
\caption{(a) Low-energy spectrum as a function of Zeeman field for a system with the same parameters as in Fig. \ref{3D_FIG2}, except $V_{BG}=520~$mV. Panels (b)-(e) show the transverse profiles of the two lowest-energy states at $x=400~$nm and Zeeman fields indicated by the red dashed lines in panel (a). Again, the wave functions swap character as $\Gamma$ increases,  indicating anti-crossing behavior. A second anti-crossing is marked by the white arrow in panel (a).}
\label{3D_FIG4}
\vspace{-3mm}
\end{figure}

The dependence of the low-energy spectrum on the Zeeman field for a system with parameters given in  Fig. \ref{3D_FIG2} is shown in Fig.\ref{3D_FIG3}(a). Note that the spectrum is particle-hole symmetric, but only the positive energy sector is shown. 
The key feature is the low energy mode that remains near zero energy from about $\Gamma \approx 0.65~$meV to $\Gamma \approx 1.15~$meV. This behavior, which is generated by the inter-band-coupling mechanism, is due to an anti-crossing between the two lowest-energy levels. To demonstrate that this is indeed the case, we calculate the  transverse profiles of the two lowest energy states at a position corresponding to the middle of the BG region and Zeeman field values on the two sides of the anti-crossing, $\Gamma\approx0.67~$meV and $\Gamma\approx1.13~$meV, respectively. Note that the transverse profile of a given state is determined by the band-components of that state, i.e. the molecular orbitals that provide the dominant contribution to the state.
The results are shown in Fig.\ref{3D_FIG3}, panels (b)-(e). The  anti-crossing is revealed by the fact that the two levels swap their transverse character, i.e. the lowest energy state at $\Gamma\approx0.67~$meV [panel (c)] becomes the second-lowest state at $\Gamma\approx1.13~$meV [panel (d)] and vice versa [see panels (b) and (e)]. This mechanism is essentially the same as the one discussed in Sec. \ref{Toy} in the context of the toy model. We remark again that in experiment, the signature of the low-energy mode will be broadened due to temperature, dissipation, and coupling to the continuum of states in the lead, which may result in the emergence of a relatively robust zero-bias conduction peak mimicking Majorana phenomenology.

To illustrate the fact that the emergence of low-energy ABSs pinned to zero energy by the inter-band coupling mechanism is quite generic, we provide another example corresponding to a larger BG gate potential, $V_{BG}=520~$mV. The results are shown in  Fig. \ref{3D_FIG4}. The spectrum shown in panel (a)  contains a low-energy mode that remains near zero energy from $\Gamma \approx 0.53~$meV to $\Gamma \approx 1.2~$ meV due to two anti-crossings. The first anti-crossing, which takes place from $\Gamma \approx 0.53$ meV to $\Gamma \approx 0.72$ meV, as marked by the red dashed lines in panel (a), is revealed by the swapping of the transverse profiles between the lowest energy levels, as shown explicitly in panels (b)-(e). 
Again, the transverse profiles correspond to the middle of the BG region and the Zeeman fields marked by the red dashed lines in panel (a). The second anti-crossing is indicated by the white arrow in Fig. \ref{3D_FIG4}(a) is revealed by a similar swapping of the transverse profile character (not shown).
This spectrum also mimics the gap opening feature predicted to occur in Majorana hybrid systems at the TQPT, simultaneously with the emergence of the Majorana mode. However, this is not a bulk gap opening, but rather a level repulsion between the lowest energy (localized) modes, which one should generically expect to occur due to inter-band coupling. We also emphasize that the only difference between the system parameters corresponding to Figs. \ref{3D_FIG3} and \ref{3D_FIG4} is the voltage applied to the BG gate. In fact, varying $V_{BG}$ -- which is exactly what is done in experiment -- provides many instances of low-energy states pinned near zero energy due to inter-band coupling. None of these states are well-separated, topologically-protected MZMs.  

\begin{figure*}[t]
\begin{center}
\includegraphics[width=\textwidth]{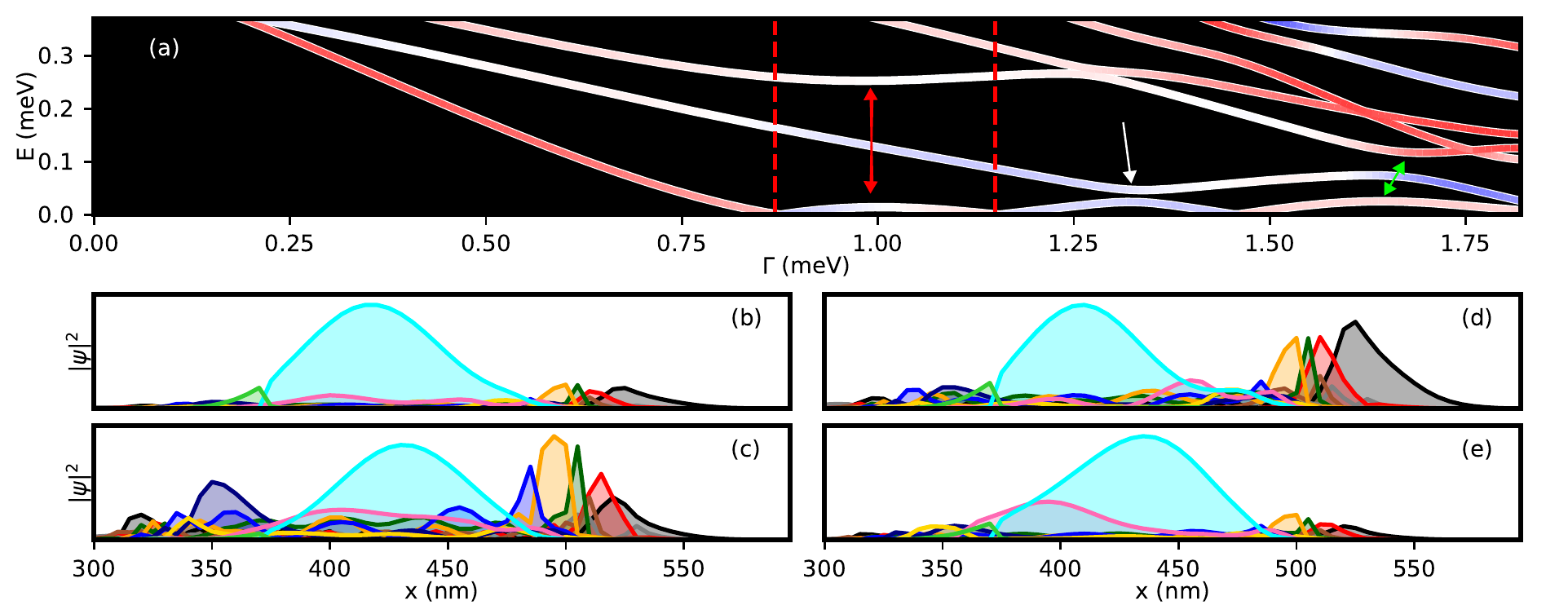}
\end{center}
\vspace{-0mm}
\caption{(a) Low-energy spectrum as a function of Zeeman field for a system with the same parameters as in Figs. \ref{3D_FIG2}-\ref{3D_FIG4}, except $V_{BG}=380~$mV. Three anti-crossings indicated by red, white, and green arrows, respectively,  pin the lowest energy mode near zero energy over a significant range of Zeeman fields.  Panels (b)-(e) show the longitudinal profiles and band composition of the first and third lowest-energy states at Zeeman fields indicated by the red dashed lines in panel (a). The color code for different band contributions is the same as in Fig. \ref{3D_FIG2}.
Note that the state in (b) has little weight in the FG region ($500$ to $600~$nm), while the state in (c) has a significant weight in both BG and FG regions. The roles change in (d) and (e), indicating that a resonance between quantum dot states associated with the BG and FG regions is responsible for this anti-crossing.}
\label{3D_FIG5}
\vspace{-3mm}
\end{figure*}

Our final example of low-energy mode pinned near zero-energy by level repulsion is shown in Fig. \ref{3D_FIG5}(a). The system parameters are the same as in Figs. \ref{3D_FIG3} and \ref{3D_FIG4}, except the BG voltage, which is $V_{BG}=380~$mV. The low-energy spectrum [panel (a)] is characterized by three anti-crossings that pin the lowest energy mode near zero energy. The anti-crossings are  indicated by red, white, and green arrows, respectively. The most obvious anti-crossing -- as revealed by a transverse profile analysis similar to those presented in Figs. \ref{3D_FIG3} and \ref{3D_FIG4} -- is the second one (indicated by the white arrow). By contrast, the first anti-crossing in Fig. \ref{3D_FIG5}(a) -- which involves the first and third energy levels -- is not clearly revealed by the transverse wave function profiles. However, it become evident if one analyses the longitudinal profile of the wave functions and their band components. The longitudinal profiles of the first and third lowest-energy modes at the Zeeman fields indicated by the red dashed lines in panel (a) are shown in panels (b)-(e). The lowest energy state shown in panel (b) is mainly composed of a single band and, more importantly, has nearly all of its spectral weight within the BG region (i.e. between $x=300~$nm and $x=500~$nm). By contrast, the third lowest energy state shown in panel (c) has significant weight in both the BG and left FG regions. Note also that this state has a larger admixture of bands, as compared to the state in panel (b), since it leaks through the BG-FG transition region, where the effective potentials vary rapidly (see Fig. \ref{3D_FIG2}) and inter-band coupling is large. For $\Gamma\approx1.15~$meV --  panels (d) and (e) -- the structures of the first and third states have (approximately) reversed, with the lowest energy state having significant weight in both the BG and FG regions, while the third lowest state being localized within the BG region. Consequently, this anti-crossing can be viewed as a resonance between two longitudinally confined, quantum dot-like, states associated with the BG and FG regions, respectively. The third anti-crossing [green arrow in panel (a)] involves a similar mechanism. We note that this coupling between two quantum dot states represent the real space counterpart of the inter-band coupling mechanism discussed in Sec. \ref{Toy}. In general, the inter-band coupling (including its real-space version -- i.e. the inter-dot coupling) acts in conjunction with the partial separation mechanism for zero-energy pinning discussed extensively in the context of single-band models. This makes the pinning near zero-energy of (topologically-trivial) ABSs a rather generic occurrence in non-homogeneous SM-SC hybrid systems with multi-band occupancy. The examples discussed in this section were obtained by changing a single experimentally-controllable parameter: the gate voltage $V_{BG}$. Varying other parameters, e.g., the gate voltage $V_{FG}$, generates similar low-energy states. The ubiquity of ABS modes pinned near zero energy by the inter-band coupling mechanism (possibly in conjunction with the partial separation mechanism) predicted by our 3D model calculations is consistent with the experimental observations on SM-SC devices with a structure similar to the setup considered here.\cite{Chen2019}  

\section{Summary and Conclusions} \label{Discussion}
In this work we have studied the emergence of low-energy ABS modes pinned near zero energy in SM-SC hybrid systems with multi-subband occupancy. We have demonstrated that the pinning of these topologically-trivial modes is due to  inter-band coupling, which occurs generically  in inhomogeneous systems. Impressive zero-energy pinning can be generated by potential inhomogeneities with rather small characteristic lengths scales  (of the order of the nanowire diameter, $100~$nm). We emphasize that this type of behavior cannot be obtained within single-band models with comparable parameters (e.g., effective mass, spin-orbit coupling, induced pairing, etc.). To get a better insight, we first illustrated the effects of the inter-band coupling mechanism using a simple multi-band toy model. 
We then 
confirmed this general picture  within a realistic 3D model that incorporates the geometric and electrostatic details of actual devices studied in the laboratory.\cite{Chen2019} The 3D calculation demonstrates that inter-band mixing occurs naturally in non-homogeneous multi-band systems due to the electrostatic-induced variation along the wire of the transverse profiles associated with different confinement-induced bands. Explicitly solving the 3D Schr\"{o}dinger-Poisson problem allows us to study  realistic device geometries without having to guess the strength of the inter-band coupling or the spacial profile of the effective electrostatic potential. We stress that within this approach there is no need to fine tune the ``intrinsic'' model parameters (e.g., effective mass, spin-orbit coupling, chemical potential, etc.) in order to pin ABSs near zero energy. Instead, one can simply tune experimentally-controllable parameters, such as, for example, the gate voltage $V_{BG}$, and identify the regimes consistent with the presence of (relatively robust) low-energy states. We emphasize that this is exactly the same protocol used in the experimental search for Majorana zero modes in SM-SC devices. 

The main implications of this study for the ongoing efforts to realize MZMs in the laboratory are fourfold. 
(1) We have shown
that the emergence of low-energy ABSs pinned near zero energy (by the inter-band coupling mechanism) is rather generic in non-homogeneous systems with multi-band occupancy. For example, many low-energy ABSs with properties similar to those illustrated in Figs. \ref{3D_FIG3}-\ref{3D_FIG5} can be obtained by sweeping the BG gate voltage within a range on the order of $1$V. 
(2) The level-repulsion generated by inter-band coupling can lead  to a rather spectacular pinning of the lowest-energy mode near zero energy in systems (or regions) characterized by very-short length scales (of the order of $100~$nm, the nanowire diameter). This demonstrates that the observation of near-zero-energy features characterized by low-amplitude energy splitting oscillations is not necessarily an indication of topological protection and well-separated MZMs. Moreover, this is  not even an indication of partial separation and quasi-Majoranas. 
(3) We have shown (see Fig. \ref{3D_FIG4}) that a level repulsion between the  lowest energy modes, which is generically induced by the inter-band coupling in the topologically trivial regime, can mimic the gap closing and re-opening feature (simultaneous with the emergence of a near zero energy mode) predicted to occur in Majorana hybrid systems at the TQPT.  This possibility has to be taken into account in the interpretation of experiments that study such features in Majorana devices.
(4) 
We identified and illustrated in Fig. \ref{FIGx} an experimental signature that could allow one to identify low-energy ABSs generated by the inter-band coupling mechanism. Specifically, any \textit{nearly} zero-bias differential conductance feature that does not exhibit particle-hole symmetry should  be attributed to the presence of (topologically-trivial) ABSs pinned near zero-energy by level repulsion, rather than  MZMs, quasi-Majoranas, or any other low-energy mode that involves (partially) separated Majorana bound states. 

Based on the results of this study, it is clear that multi-band physics significantly complicates the interpretation of any experiment involving SM-SC hybrid structures, in particular charge tunneling measurements. An obvious way to reduce the importance of inter-band coupling is to reduce the diameter of the wire as much as possible, without inducing disorder. This will increase the energy spacing between bands and, therefore, reduce the importance of inter-band coupling. On the other hand, large diameter nanowires tend to approach the regime in which many confinement-induced sub-bands cluster near the chemical potential generating large inter-band couplings that control the low energy physics of the system. A second path toward reducing inter-band coupling is to use negative (rather than positive) voltages on the back gates. On the one hand, this reduces the number of  occupied bands. On the other hand, it increases the inter-band energy spacing for two reasons; (1) the lowest-energy confinement-induced conduction bands tend to have larger energy spacing due to a lower effective mass (as compared to the high-energy bands) and (2) the negative voltage pushes the wave functions towards the SM-SC interface, increasing the confinement and, consequently, the inter-band spacing.
Note that the negative gate voltage can also reduce the inter-band superconducting pairing. 
Finally, our findings highlight the importance of inhomogeneous effective potentials in generating low-energy ABSs. While inhomogeneous potentials have been previously shown to induce low energy topologically-trivial ABSs, this study reveals that  in multi-band systems the collapse and pinning of ABSs to zero energy can take place  even when the characteristic length scale of the potential non-uniformity is on the order of $100~$nm. We stress that the electrostatic gradients between different regions of the wire need to be as sharp as possible to reduce inter-band coupling. This problem is the multi-band generalization of the sharp versus smooth confinement discussed extensively within the context of single-band toy models.

\begin{acknowledgments}
T.D.S. acknowledges  NSF DMR-1414683. S.M.F. acknowledges NSF DMR-1743972, NSF PIRE-1743717, ONR and ARO.
\end{acknowledgments}

\appendix

\section{Inter-band spin-orbit coupling} \label{App1}
Consider a two subband model with longitudinal spin orbit coupling strength $\alpha$. Suppose initially that the potential is homogeneous such that there is no inter-subband coupling. The portion of the spin-orbit coupling matrix coupling lattice site $i$ to site $i+1$ is then given by
\begin{equation}
(H_{\alpha})_{i,i+1} = 
\begin{bmatrix}
0 & \alpha & 0 & 0 \\
-\alpha & 0 & 0 & 0 \\
0 & 0 & 0 & \alpha \\
0 & 0 & -\alpha & 0 
\end{bmatrix},
\end{equation}
where the first and last two columns describe the coupling for the first and second subbands, respectively. Suppose we now apply a potential such that the basis states of site $i+1$ are slightly rotated in the Hilbert space. Let us denote these basis states by
\begin{eqnarray}
\left|+,\sigma\right>_{i+1} = \sqrt{ \frac{n-1}{n}} \left|1,\sigma\right>_{i+1} + \sqrt{ \frac{1}{n}} \left|2,\sigma\right>_{i+1}, \\
\left|-,\sigma\right>_{i+1} = -\sqrt{ \frac{1}{n}} \left|1,\sigma\right>_{i+1} + \sqrt{ \frac{n-1}{n}} \left|2,\sigma\right>_{i+1},
\end{eqnarray}
where $n\geq 1$, $\sigma$ and $1$ in $\left|1,\sigma\right>_{i+1}$ denote the spin label and first band in the original basis, and the ($i+1$) subscript indicates the $i+1$ site. These eigenstates are orthonormal as can be easily check. What is the new spin orbit coupling matrix coupling the basis states of sites $i$ and $i+1$? We simply calculate the matrix elements $ {}_{i}\left<n,\sigma\left|H_\alpha\right| \pm,\sigma^\prime\right>_{i+1}$, where n =1,2. The new matrix becomes
\begin{equation}
(\widetilde{H}_{\alpha})_{i,i+1} = \frac{\alpha}{\sqrt{n}} 
\begin{bmatrix}
0 & \sqrt{ n-1} & 0 & - 1 \\
-\sqrt{ n-1} & 0 & 1 & 0 \\
0 & 1 & 0 & \sqrt{ n-1} \\
-1 & 0 & -\sqrt{ n-1} & 0 
\end{bmatrix}
\end{equation} 
\vspace{1mm}
This simple examples shows why we assume $\widetilde{\alpha}_{i}^{\alpha\beta} = - \widetilde{\alpha}_{i}^{\beta\alpha}$ in the inhomogeneous potential cases of Sec \ref{Inhomo}. 

\vspace{2mm}
While this type of coupling is perfectly fine in the case of an inhomogeneous potential, it is not physical in the case of a homogeneous potential with $\Delta_{\alpha\beta}=\Delta_{\beta\alpha}$. This is because the anti-symmetry of the interband spin-orbit coupling breaks inversion symmetry (i.e. $\sigma_x U_I^\dagger H U_I \sigma_x \neq H$, where $U_I$ inverts the sites through the center of the wire and $\sigma_x$ is the Pauli spin flip operator). For this reason, we can not include a term $\widetilde{\alpha}_{i}^{\alpha\beta} = - \widetilde{\alpha}_{i}^{\beta\alpha}$ in the cases of homogeneous potential discussed in Sec \ref{Homo}. Rather we employ transverse interband spin orbit coupling $(\alpha_T)^{\alpha\beta}_i$, which couples the various subsbands without breaking inversion symmetry.

\section{Majorana Representation} \label{App2}
Recall that, within the BdG formalism, each positive energy state $\psi_n$ with energy $E_n$ has a corresponding negative energy state $\psi_n^\prime$ with energy $-E_n$. We decompose the lowest energy mode represented by $\psi_1$ and its negative energy partner $\psi_1^\prime$ into two Majorana modes, $\chi_1$ and $\chi_2$, given by 
\begin{align}
\chi_1 = \frac{1}{\sqrt{2}}\left[ \psi_1 + \psi_1^\prime\right], 
\end{align}
\begin{align}
\chi_2 = \frac{i}{\sqrt{2}}\left[ \psi_1 - \psi_1^\prime\right].
\end{align}
The Majorana modes, $\chi_1$ and $\chi_2$, are generically not eigenstates of the BdG Hamiltonian, except for $E_1 = 0$. However, the degree of overlap between the two Majorana modes provides useful information regarding the robustness of the low-energy model to changes in the system parameters such as the applied magnetic field.\cite{Moore2018} In particular, small overlap between $\chi_1$ and $\chi_2$ indicates a low-energy state that remains near energy for a large range of magnetic field. These states are then topological MZMs or so called ps-ABS if the toplogical condition is not met globally for the entire system. In contrast, large overlap of $\chi_1$ and $\chi_2$ indicates that the low-energy states are quite sensitive to the system parameters, as is the case for trivial ABS. Generically, these states should not be expected to stick to zero energy. However, in the multi-band case, inter-band level repulsion can keep these trivial ABS near zero energy even though the Majorana modes of which they're composed are significantly overlapping (see main text).

\end{document}